\documentclass[11pt,a4paper]{article}
\usepackage{jcappub}
\usepackage{hyperref}
\usepackage[english]{babel}
\usepackage{graphicx}
\usepackage{bm}
\usepackage{amssymb}
\usepackage{amsmath}
\usepackage{amsfonts}
\usepackage{footnote}   % for footnotes in tables

\newcommand{\grado}{^{\circ}}
\def\msun{\mbox{$M_\odot$}}

\title{Dark matter searches with Cherenkov telescopes: nearby dwarf galaxies or local galaxy 
clusters?}

\author[a,b,c]{Miguel A. S\'anchez-Conde,}
\author[d]{Mirco Cannoni,}
\author[e]{Fabio Zandanel,}
\author[d]{Mario E. G\'omez}
\author[e]{and Francisco Prada}

\affiliation[a]{SLAC National Laboratory and Kavli Institute for Particle Astrophysics and Cosmology, 2575 Sand Hill Road, Menlo Park, CA 94025, USA}

\affiliation[b]{Instituto de Astrofisica de Canarias, E-38205 La Laguna, Tenerife,
Spain}

\affiliation[c]{Dpto. Astrof\'isica, Universidad de La Laguna (ULL),
E-38205 La Laguna, Tenerife, Spain}

\affiliation[d]{ Dpto. F\'isica Aplicada, Facultad de Ciencias
Experimentales, Universidad de Huelva, 21071 Huelva, Spain}

\affiliation[e]{Instituto de Astrof\'isica de Andaluc\'ia (CSIC), E-18008, Granada,
Spain}

\abstract{
In this paper, we compare dwarf galaxies and galaxy clusters in order to elucidate which object class is the best target for gamma-ray DM searches with imaging atmospheric Cherenkov telescopes (IACTs). 
We have built a mixed dwarfs+clusters sample containing some of the most promising nearby dwarf galaxies (Draco, Ursa Minor, Wilman~1 and Segue~1) and local galaxy clusters (Perseus, Coma, Ophiuchus, Virgo, Fornax, NGC~5813 and NGC~5846), and then compute their DM annihilation flux profiles by making use of the latest modeling of their DM density profiles. We also include in our calculations the effect of DM substructure. 
Willman~1 appears as the best candidate in the sample. However, its mass modeling is still rather uncertain, so probably other candidates with less uncertainties and quite similar fluxes, namely Ursa Minor and Segue~1, might be better options. As for galaxy clusters, Virgo represents the one with the highest flux. However, its large spatial extension can be a serious handicap for IACT observations and posterior data analysis. Yet, other local galaxy cluster candidates with more moderate emission regions, such as Perseus, may represent good alternatives. 
After comparing dwarfs and clusters, we found that the former exhibit annihilation flux profiles that, at the center, are roughly one order of magnitude higher than those of clusters, although galaxy clusters can yield similar, or even higher, integrated fluxes for the whole object once substructure is taken into account. 
Even when any of these objects are strictly point-like according to the properties of their annihilation signals, we conclude that dwarf galaxies are best suited for observational strategies based on the search of point-like sources, while galaxy clusters represent best targets for analyses that can deal with rather extended emissions. 
Finally, we study the detection prospects for present and future IACTs in the framework of the constrained minimal supersymmetric standard model. We find that the level of 
the annihilation flux from these targets is below the sensitivities of current IACTs and the 
future CTA. 
}

\keywords{Dark matter, supersymmetry, dwarf galaxies, cluster of galaxies, gamma rays, Cherenkov 
telescopes}

\arxivnumber{1104.3530}

\begin{document}
\maketitle
\flushbottom

\section{Introduction}

During the last century, many astrophysical observations on different scales seemed 
to point to the fact that the luminous matter in the Universe is just a tiny fraction of its total 
content. Effectively, there exists  strong evidence for believing that most of the matter in our Universe 
is dark. This dark matter (DM) has not been directly detected in the laboratory yet, but its gravitational 
effects have been observed on all spatial scales, from the inner kiloparsecs of galaxies out to Mpc and 
cosmological scales. One of the first steps in the DM paradigm was taken by F. Zwicky in the 1930s 
to explain the velocity dispersion in galaxy clusters. Today, the most conclusive observations in this 
sense come from the rotational speeds of galaxies, the orbital velocities of galaxies within clusters, 
gravitational lensing, satellite kinematics, the cosmic microwave background, the light element abundances 
and large scale structure \cite{review}.

However, we still do not know what the 
DM  is made of. Physics beyond the Standard Model of Particle Physics needs to be invoked, where 
good non-baryonic DM candidates arise to fulfill all the cosmological requirements.  In the Lambda 
Cold Dark Matter ($\Lambda$CDM) paradigm, 
around 23\% of the Universe consists of non-baryonic DM \cite{wmap07}. A plethora of possible 
DM candidates have already been proposed. Axion-like particles, proposed for different reasons 
in an extension of the Standard Model of particle physics, represent one of the most popular candidates 
\cite{review2,masc_axions}. Ordinary massive neutrinos are too light to be cosmologically significant, 
though sterile neutrinos remain a possibility. Other candidates include primordial black holes, 
non-thermal WIMPzillas, and Kaluza-Klein particles (see  Refs.~\cite{bertone07,bergstrom10} for a 
recent  and detailed picture). 
The neutralino is probably the most studied candidate and it arises in supersymmetric (SUSY) extension 
of the Standard Model of particle physics. In particular,  in the minimal supersymmetric standard model (MSSM)
R-parity is conserved and the the lightest SUSY particle is one of the neutralinos, which is a stable Majorana particle 
with a relic density compatible with WMAP bounds. In this work we consider the lightest neutralino to be the 
DM particle. Being the neutralino its own antiparticle, it annihilates when interacting with other 
neutralinos. This fact is crucial for detectability purposes, 
as one of the products of these annihilations are predicted to be gamma-rays, 
whose specific energy will vary according to the chosen particle physics model, but that is expected to 
lie in the energy range covered by the current imaging atmospheric Cherenkov telescopes (IACTs). The 
imaging atmospheric Cherenkov technique, first pioneered by the Whipple collaboration, currently leads above 100 GeV by HESS 
\cite{hess}, MAGIC \cite{magic} and VERITAS \cite{veritas}. Also the NASA Fermi satellite \cite{fermi} 
is playing a major role in the exploration of the gamma-ray energy regime, namely between 20 MeV up to 
$\sim$300 GeV. This study, however,  will mainly focus on DM searches with IACTs.%, although we will discuss possible synergies between IACTs and Fermi.

The DM annihilation flux is proportional to the annihilation rate, which is proportional 
to the squared DM density. This means that the best places to look for DM will be those 
with the highest DM concentrations. Distance is also very important, since highly DM-dominated systems that 
are located too far from us will yield too low DM annihilation fluxes at Earth. Keeping both considerations 
in mind, IACT efforts have  focused on the Galactic Center \cite{hess04,magic06, hess06GC} and dwarf 
galaxy satellites of the Milky Way \cite{evans,StrigariPRD}. Nearby galaxy clusters may also represent 
very suitable targets, given their large DM content; indeed, they may yield similar gamma-ray fluxes 
despite their distance (see e.g.~\cite{pinzke}). At present, many of these objects have already been 
observed in $\gamma$-rays. Unfortunately, no clear signal from 
DM annihilation has been found yet (at least unequivocally; for example, see  \cite{hooper_review} for some hints of 
detection). In the last few years, most  attention have been devoted to the search for neutralino 
annihilations in nearby dwarf galaxies 
\cite{MAGICdraco,SagHESS,whippleobs,MAGICwillman1,veritasdwarfs,hess09CMa,MAGICsegue,HESScarinasculptor} 
rather than in massive nearby galaxy clusters, although up to now no one has studied in detail which kind 
of object class would be preferable for DM searches. 

The main aim of this study is to compare both 
the expected DM annihilation fluxes from the most promising dwarfs and galaxy clusters 
in order to elucidate which object class is most suitable for gamma-ray DM 
searches with present and future IACTs. We make use of the latest modeling of  DM density 
profiles calculated from the latest available observational data. We also carefully 
include the effect of DM substructure, taking as reference the procedure described 
in Ref.~\cite{kkk}. 
On the particle physics side, we do include internal bremsstrahlung in the computation of the number 
of photons per annihilation (see, for example, \cite{IB}). This effect might significantly enhance the DM annihilation 
flux for specific models of the allowed parameter space \cite{cannoni}. 
We do not consider, however, models with Sommerfeld enhancement, which may provide an additional boost to the flux \cite{sommerfeld}\footnote{This effect depends on the mass of the DM particle and its velocity, which means that it is necessary to know in detail the velocity distribution of the DM particles inside the objects. This is beyond the scope of this paper. Furthermore, the importance of the Sommerfeld effect is very sensible to e.g. little variations of the considered DM particle mass or other slight changes in the involved parameters (as shown e.g. in Ref.~\cite{pieri09sommerf}). Therefore, we decided not to include this effect and to keep conservative in deriving our detection limits. In any case, we note that according to recent estimates, Sommerfeld enhancements are not probably larger than $\sim$5 for clusters \cite{pinzke11}.}.

We furthermore perform a detailed analysis of the expected gamma-ray DM annihilation flux from each object 
by defining and calculating some specific quantities that were carefully chosen taking into account both the
instrumental and observational properties of IACTs. These new parameters will contribute with relevant 
information to our understanding of the particular characteristics that a gamma-ray flux with a DM-annihilating 
origin may necessarily exhibit. 
We select our sample according to i) previous estimates of the DM annihilation flux, ii) distance, iii) an acceptable knowledge of the DM content, and iv) abundant literature. With 
all these considerations in mind, we finally focused our efforts on four dwarf galaxy satellites of the Milky
 Way (Draco, Ursa Minor, Willman~1, and Segue~1), and seven galaxy clusters (Perseus, Coma, Ophiuchus, 
 Virgo, Fornax, NGC~5813, and NGC~5846). We refer the reader to sections \ref{sec:dwarfs} and \ref{sec:clusters} 
 respectively for further details on the sample selection.

This article is organized as follows. In section \ref{sec:flux} we discuss the factors involved in the computation of the 
gamma-ray DM annihilation flux: the astrophysical factor in section \ref{sec:astrofactor} (with the definition of some reference values in~\ref{subsub}), and the particle physics factor in \ref{sec:fsusy}. In section \ref{sec:dwarfs} we perform a detailed study of the DM annihilation flux for our sample of dwarf galaxies, discussing the best 
observational strategies to be followed for a successful search. A similar strategy is followed for galaxy clusters in 
section \ref{sec:clusters}, this time including substructure as well. We finally compare the DM annihilation 
prospects for dwarfs and clusters in section \ref{sec:dwarfsvsclusters} and state our conclusions in section \ref{sec:conclusions}.

\section{Flux of gamma rays from DM annihilation}  \label{sec:flux}

Why $\gamma$-rays and not other wavelengths? The key point is the energy scale of the annihilation 
products, which is determined by the mass of the DM particles. Since  preferred DM candidates like the neutralino are expected to 
have masses in the $\sim$GeV--TeV range\footnote{
The evidence of an annual modulation in the rate of elastic
WIMP-nucleus scattering claimed by DAMA \cite{dama} and 
recently by CoGENT \cite{cogent} favours a light WIMP with mass around 10 GeV that is not 
an ideal candidate for gamma ray detection with IACT. The results of other 
experiments are in contrast with this conclusion, thus at present 
there is no clear indication in favour or against a light WIMP~\cite{cannoni11}.}, DM searches are specially performed in the 
$\gamma$-ray energy band.  
%Furthermore, DM searches based on gamma rays have some advantages with respect to other indirect detection techniques (e.g., \cite{hooper_review}): they travel essentially unimpeded (at least in the Milky Way's ``cosmological'' neighborhood), since they are unaffected by astrophysical magnetic fields, contrary to what happens for charged particles (electrons, positrons, etc.). In addition, they are not attenuated for small cosmological distances, in this way retaining their spectral information, i.e. the observed spectrum at Earth will be identical to the DM annihilation spectrum, which  depends only on the WIMP considered. 
However, we stress that 
multi-wavelength studies could be of considerable importance in constraining  DM models and therefore should also 
be taken into account in order to reach a consistent general picture (e.g., Refs.~\cite{mapt,colafrancesco06,colafrancesco10}).

In this study we will make the assumption that the neutralino $\chi$ is the main component of the DM\footnote{Note that the total amount of DM in the Universe could not be constituted by a single class of DM particle. 
Indeed, we already know that standard neutrinos contribute to DM, although they cannot account for all of it. A discussion on the detection prospects of a sub-dominant density component of DM can be found, for example,
 in \cite{duda03}.}. The expected total number of continuum $\gamma$-ray photons 
received per unit time and per unit area, above the energy threshold E$_{th}$ of the telescope, when observing 
at a given direction $\Psi_0$ relative to the centre of the DM halo is given by:
\begin{equation}
F(E_{\gamma}>E_{\rm th}, \Psi_0)=J(\Psi_0)\times{f_{SUSY}(E_{\gamma}>E_{\rm th})}.
\label{eq1}
\end{equation}
The factor $f_{SUSY}$ incorporates all the particle physics, whereas all the astrophysical considerations 
are included in $J(\Psi_0)$. We discuss both factors in the next two subsections.

\subsection{$J(\Psi_0)$: the astrophysical flux factor}   \label{sec:astrofactor}

\noindent $J(\Psi_0)$ in eq.(\ref{eq1}) accounts for the DM distribution, the geometry of the problem, and 
also the instrumental effects induced by the IACT; i.e.,
\begin{equation}
J(\Psi_0)=\frac{1}{4\pi} \int U(\Psi_0)B(\Omega)d\Omega,
\label{eq1c}
\end{equation}
where $B(\Omega)d\Omega$ represents the beam smearing of the
telescope, commonly known as the point spread function (PSF). The PSF
can be well approximated by a Gaussian:
\begin{equation}
B(\Omega) d\Omega  = \frac{1}{\sqrt{2\pi \sigma_t^2}} \exp\left[ -\frac{\theta^2}{2\sigma_t^2}\right] 
\sin\theta ~d\theta ~d\phi,
\label{eq3}
\end{equation}
with $\sigma_t$ the angular resolution of the IACT. The $U(\Psi_0)$ factor of eq.(\ref{eq1c}) represents the 
integral of the line of sight (l.o.s.) of the DM density squared along the
direction of observation $\Psi_0$:
\begin{equation}
U(\Psi_0) = \int_{l.o.s.} \rho_{DM}^2(r)~d\lambda =
\int_{\lambda_{min}}^{\lambda_{max}} \rho_{DM}^2[r(\lambda)]~d\lambda .
\label{eq2}
\end{equation}
Here, $\rho_{DM}$ is the DM density profile and $r$ represents the 
galactocentric distance, related to the distance $\lambda$ to the Earth by:
\begin{equation}
r = \sqrt{\lambda^2+R_{\odot}^2-2~\lambda~R_{\odot} \cos \Psi},
\label{eq1a}
\end{equation}
where $R_{\odot}$ is the distance from the Earth to the
centre of the halo, and $\Psi$ is related to the angles $\Psi_0$, 
$\theta$ and $\phi$ by the relation $\cos \Psi = \cos \Psi_0 \cos
\theta + \sin \Psi_0 \sin \theta \cos \phi$. The lower and
upper limits $\lambda_{min}$ and $\lambda_{max}$ in the
l.o.s.\ integration are given by $R_{\odot}\cos \Psi \pm
\sqrt{r_t^2-R_{\odot}^2\sin^2 \Psi}$, where $r_t$ is the radius of the object under consideration (e.g., the 
tidal radius in case of dwarfs).\\

One crucial aspect in the calculation of the astrophysical factor concerns the modeling of the DM distribution. 
Current $N$-body cosmological simulations suggest the existence of a universal DM density profile, with the same shape 
for all masses and epochs \cite{nfw96}. In Ref.~\cite{kravtsov}, authors proposed a general parameterization for the 
DM halo density in order to agglutinate most of the fitting formulae that can be found in the literature:
\begin{equation}
\rho(r) = \frac{\rho_0}{\left(\frac{r}{r_s}\right)^\gamma \left[1 +\left(\frac{r}{r_s}\right)^\alpha\right]^
{\frac{\beta - \gamma}{\alpha}}}, 
\label{densityprofile}
\end{equation}
where $\rho_0$ and $r_s$ represent a characteristic density and a scale radius respectively. The 
Navarro-Frenk-White (hereafter NFW), with ($\alpha$,$\beta$,$\gamma$) = (1,3,1), is by far the most widely DM 
density profile used in the literature \cite{nfw97}. Other studies seem to point to inner slopes of 1.2 \cite{diemandcusps}. Recently, also the so-called Einasto DM density profile 
is often used, motivated by the fact that state-of-the-art $N$-body cosmological simulations \cite{stadel09,navarro10} 
seem to point to a less steep DM density profile in the central regions of CDM halos, which in addition  
shows better agreement with observations \cite{navarro04,merritt06,gao08}:
\begin{equation}
\rho(r) = \rho_s\exp\left[\frac{-2}{\alpha}~\left(\left(\frac{r}{r_s}\right)^{\alpha}-1\right)\right].
\label{eq:einasto}
\end{equation}

The main controversy between different groups pertains to the slope of this profile in the innermost regions of 
galaxies and galaxy clusters, which are obviously the most conflictive regions to be simulated correctly. Uncertainties arise also from the effects of the adiabatic baryonic compression of the DM near the halo center \cite{pradaPRL}, 
typically not considered in $N$-body simulations, as well as from the possible effect of central black holes (e.g., \cite{gnedin&primack,diemand05}).

The PSF of the telescope, on the other hand, plays 
a crucial role in the correct interpretation of the observational data~\cite{masc}. The present generation of IACTs have PSF$\sim$0.1$\grado$ 
(i.e., $\sigma_t$=0.1$\grado$ in eq.(\ref{eq3})), which means that any gamma source with an extension smaller 
than this value will appear as point-like to the telescope. For our purposes, it is important to understand 
that point-like, rather than extended, sources are more readily observable by IACTs\footnote{IACT sensitivity is approximately linearly proportional to the source extension. This means that the detectability of a gamma-ray source is proportional to its luminosity divided by its size. This fact clearly makes point-like sources easier targets for IACTs.}. This is particularly 
true for single telescopes, for which stereoscopic analysis is not possible. However, the DM gamma signal is 
expected to be extended, since it follows the DM distribution; hence it is clear that a good DM-oriented 
observational strategy should be able to find a compromise between these crucial instrumental/observational 
aspects and the peculiarities of the DM gamma emission in the most interesting astrophysical targets. 

\subsubsection{Reference values of the astrophysical factor}
\label{subsub}

We will now define some quantities that will become extremely helpful throughout the rest of the paper.\\

J$_T$ is the total astrophysical factor 
obtained integrating over the total angular extension of the object:
\begin{equation} 
\text{J$_{T}$}=\frac{1}{4\pi D^2} \int_{V} \rho_{DM}^2(r)~dV,
\end{equation}
$D$ being the distance from the Earth to the center of the DM halo and $r$ the galactocentric distance 
inside it. Note that this expression no longer depends on the PSF. 

Quantities related to  the {\it observed} (i.e., taking into account the PSF) flux profile up to a specific radius/angle are calculated with:
\begin{equation}
\text{J}_{\psi'}= \frac{1}{4\pi}\times 2\pi \times \int_{0}^{\psi'}J(\Psi_0)~\sin(\Psi_0)~d\Psi_0
\label{eq:integprof}
\end{equation}
where $J(\Psi_0)$ is given by eq.~~(\ref{eq1c}). In particular:
\begin{itemize}
\item %[-]
$\psi' = \psi_{90}$ and r$_{90}$ are, respectively, the angle and the radius that identify the region
from which 90\% of the gamma emission is expected to come. $\text{J}_{90}$
indicates the corresponding astrophysical factor.
\item %[-]
$\psi' = 0.1^{\circ}$ and r$_{01}$ refers to the inner $0.1^{\circ}$ of the halo,
and J$_{01}$ is the corresponding astrophysical factor.
\item %[-]
$\psi'=\psi_{r_s}$ is the angle subtended by the scale radius r$_s$, and J$_{r_s}$ is the astrophysical factor
associated to the region inside r$_s$;
\item
We further define 
$Rank_{01}$ and $Rank_{90}$  which
order the candidates
according to the values of
J$_{01}$ and $\text{J}_{90}$, respectively.
\end{itemize}

We chose quantities related to 
90\% of the annihilation flux because this is typically the fraction of the total flux that comes 
from the region within r$_s$ for an NFW profile. Therefore, it is probably better to plan 
observational strategies focused on detecting the flux from this smaller area rather than 
from the total extension of the source\footnote{In the latter case, we will be introducing a lot of 
extra background without significantly increasing the annihilation signal. In dwarfs, for 
instance, r$_s$ typically represents less than 10\% of the total radius.}. Furthermore, point-like 
sources are more readily observable by present IACTs, as already discussed above. This information is codified in J$_{01}$ and r$_{01}$, which were specially selected by comparison with the typical IACT PSF $\sim$ 0.1$^{\circ}$. In particular, the 
larger  values of J$_{01}$/J$_T$ give more point-like  objects that are more easily detected.

\subsection{The particle physics factor, $f_{SUSY}$, in CMSSM}
\label{sec:fsusy}

As already pointed out, we consider $R$-parity conserving 
SUSY models such that the LSP, the lightest neutralino, is the main dark matter component. 
The mechanisms producing photons in neutralino annihilation are well known. The dominant 
contribution is typically constituted by {\it secondary photons} coming from the hadronisation 
and decay of the annihilation products~\cite{gammas}. The energy spectrum is continuous and 
decreasing towards $m_{\chi}$.
At the one loop level neutralinos annihilate into photons through the processes~\cite{lines}
$\chi \chi \rightarrow \gamma \gamma$ and
$\chi \chi \rightarrow Z \gamma$.
Outgoing photons are almost monochromatic ({\it lines}) with energies
$E_\gamma\sim m_\chi$ and 
$E_\gamma\sim m_\chi- {m_Z^2 /{4 m_\chi}}$,
respectively. Although these gammas would give a very clear signal, the cross
section is ${\mathcal O}$($\alpha^4$ ), thus very small. 
Finally, {\it internal bremsstrahlung} (IB) \cite{IB} which consists on the emission of additional photons  
from neutralino pair annihilation into charged particles:  
$\chi\chi \to X\bar{X}\gamma$, $X$ being a charged lepton or a $W$ boson.  
The cross section is ${\mathcal O} $($\alpha^3$ ), thus in principle its contribution is between the tree-level secondaries and 
the loop-level monochromatic lines. However, its relevance strongly depends on the SUSY spectrum \cite{cannoni}. 
$f_{SUSY}$ is thus given by:
\begin{eqnarray}
f_{SUSY} &=&f_{cont}+f_{lines},\\
f_{cont}&=&\left( \sum_f \int_{E_{th}}^{m_{\chi}}
\frac{dN^{f}_{\gamma}}{dE_{\gamma}}dE_{\gamma} \right)
\frac{\langle \sigma_{\chi\chi} \,v\rangle} {2 m_{\chi}^2} = f_{sec} + f_{IB}, \\
f_{lines} &=& 2 \frac{\langle \sigma_{\gamma\gamma} \,v \rangle} {2 m_{\chi}^2} +
\frac{\langle \sigma_{Z\gamma} \,v\rangle}{2 m_{\chi}^2}.
\label{fsusy}
\end{eqnarray}
${dN^{f}_{\gamma}}/{dE_{\gamma}}$ is
the differential yield of photons per annihilation to
the final state $f$, thus $n_{\gamma} (E_{\gamma} > E_{th})$ is the total number of photons per
annihilation with energy greater than the threshold energy. $\langle \sigma_{\chi\chi}
\,v\rangle$ is the thermal averaged total neutralino annihilation cross section,  
$\langle \sigma_{\gamma\gamma} \,v \rangle$ and 
$\langle \sigma_{Z\gamma} \,v\rangle$ the cross sections for annihilation into
lines and $m_{\chi}$ the neutralino mass.

The soft potential of the general MSSM at the weak scale contains more than a
hundred  free parameters: this large number
is drastically reduced by adding further 
theoretical assumptions at the energy scale
where the gauge couplings constants of the standard model unify (GUT scale).
The benchmark theory with a low number of free parameters,
which is the subject of the majority of  phenomenological studies, 
is the constrained minimal supersymmetric standard model (CMSSM)~\cite{susylag}.
The theory at the weak scale is 
determined by  four parameters: 
the common scalar mass $m_0$, the common gauginos mass $m_{1/2}$,  
the common trilinear couplings $A_0$ assigned at the GUT scale,
and the ratio of the Higgs vacuum expectation values,
$\tan\beta$. The sign of $\mu$, the Higgs mixing term,
is left undetermined: we consider in the following the case
$\mu >0$.
We select SUSY models such that the relic abundance of the neutralino is inside 
the cosmologically favoured WMAP interval~\cite{WMAP}, $0.09 <\Omega_\chi h^2 < 0.13$ at $3\sigma$.
In addition we impose the LEP bounds on the masses of the light Higgs and the chargino,
$m_h >114$ GeV and $m_{\chi^+} > 103.5$ GeV~\footnote{
In the CMSSM and other SUSY models with gaugino mass unification,
the mass of the charginos and the mass of the neutralinos are strictly correlated.
The bound on the lightest chargino mass does not allow for a light neutralino with mass
below $\simeq$ 50 GeV.}, and the constraints from $b\to s\gamma$.

For fixed values of $\tan\beta$ and $A_0$,
only narrow regions of the CMSSM  parameter space in the ($m_{1/2}$, $m_0$) plane
pass the phenomenological constraints: 
1) the stau coannihilation region where the next
lightest SUSY particle, the scalar partner
of the tau fermion, is nearly degenerate in mass with the lightest neutralino;
2) the Higgs funnel region where the neutralino mass is nearly half of the
pseudo-scalar Higgs boson $A$; 3) the focus-point region at large $m_0$ where the 
neutralino has a large higgsino component.
\begin{table}[t]
\resizebox{\textwidth}{!}{
\begin{tabular}{ccccccccc}
\hline
%\\
Model & $\tan\beta$ & $m_{0}$   & $m_{1/2}$  & $A_{0}$  & $m_{\tilde{\chi}}$ &
$\Omega_\chi h^2$ & $\langle \sigma_{\chi\chi} v \rangle$   &
$f_{SUSY}$ \\
      &         & (GeV)& (GeV) & (GeV) & (GeV) &    &(cm$^{3}$ s$^{-1}$) &
(GeV$^{-2}$cm$^{3}$s$^{-1}$)  \\
\hline
%\hline

A & $18$ & $127$ & $459$ & $-135$ & 187.6 & 0.092 & $0.03\times 10^{-26}$    & $0.1 \times 10^{-32}$\\
%\hline
B & $52$ & $982$ & $1377$ & $725$ & 597.6 & 0.092 & $2.6\times 10^{-26}$    & $0.72 \times 10^{-32}$ \\
%\hline
C & $17$ & $2200$ & $430$ & $805$ & 162.8 & 0.098 & $2.2\times 10^{-26}$  & 0.12 $\times 10^{-32}$ \\
%\hline
D & $51$ & $8940$ & $2218$ & $-4221$ & 918.2 & 0.099  & $1.2\times 10^{-26}$   & $0.32 \times 10^{-32}$ \\
\hline
\end{tabular}}
\caption{Reference models in the CMSSM with $\mu>0$ and the corresponding relevant physical quantities 
for $\gamma$-ray production and detection. The value of the particle physics factor $f_{SUSY}$ is 
obtained integrating on photon energies above the threshold $E_{th} =100 $ GeV.}
\label{tab}
\end{table}

\begin{figure}[t!]
\centering
\includegraphics*[scale=0.4]{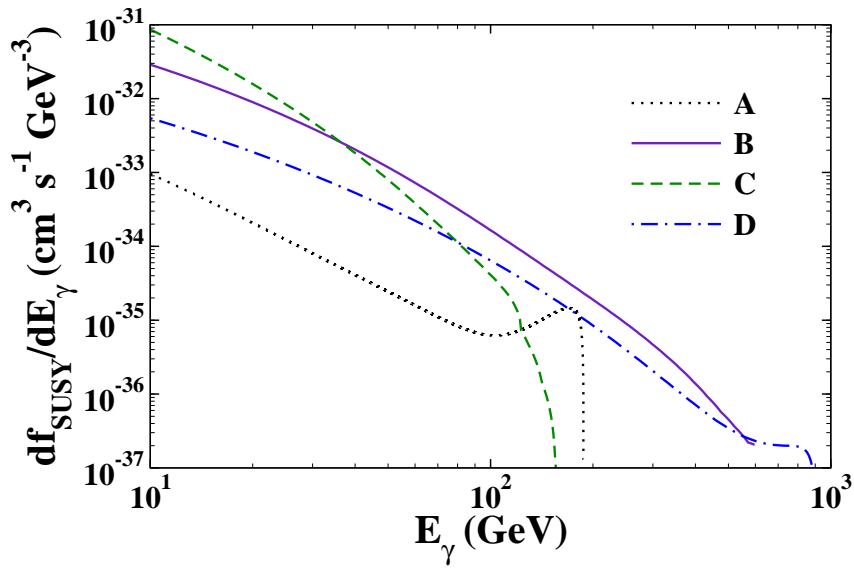}
\caption{Differential particle physics factor as a function of $\gamma$-ray energy for the models reported in table~\ref{tab}. See text for further details and discussion.} 
\label{fsusydif}
\end{figure}

In table~\ref{tab}  we provide some reference models chosen from 
an exhaustive scan of the CMSSM parameter space in Ref.~\cite{cannoni}.
Model A is in the stau coannihilation region, 
Model B is in the funnel region and Models C and D are in the focus point region.
The numerical values for the neutralino mass, the relic density, the total annihilation cross
section and $f_{SUSY}$ with $E_{th} =100 $ GeV were obtained using 
the code \textsf{DarkSusy 5.0.5}~\cite{darksusy}.

The most optimistic value of $f_{SUSY}$, 
with a typical IACT threshold energy $E_{th} =100$ GeV,
is $\simeq 10^{-32}$ GeV$^{-2}$ cm$^{3}$ s$^{-1}$ and it is 
found in Model B. 
As explained above, in this case $m_A \simeq 2m_\chi$.
The large annihilation cross section 
is determined by the diagram with $s$-channel resonant pseudo-scalar Higgs boson propagator.
The branching ratios of the final states from $A$ decay
are $\sim 87\%$ in $b\bar{b}$ and $\sim 13\%$ in $\tau^+ \tau^-$
that result in a large amount of gamma in the subsequent hadronization.

Model C and D  predict values of $f_{SUSY}$ of the same order of magnitude but smaller than 
Model B. In these cases the main annihilation final states are $W^+ W^-$ and $ZZ$
due to the larger higgsinos component in the neutralino that enhances the coupling to the gauge 
bosons. 

In Model A the only relevant final states are $b\bar{b}$ and $\tau^+ \tau^-$ 
(both with fraction around $50\%$) and the neutralino is bino-like as in 
Model B. In this case the resonant condition $m_A \simeq 2m_\chi$ is not satisfied thus the cross
section, as can be seen in table~\ref{tab}, is two orders of magnitude smaller. 
In this model, on the other hand, the mass of the lightest stau is close to the neutralino mass,
$m_{\tilde{\tau}} \gtrsim m_{\chi}$. The diagram with $t$-channel stau exchange, hence,  
is not suppressed with respect to diagrams with Higgs boson $s$-channel exchange.
Furthermore, due to this mass degeneracy, the diagram with a photon line attached to 
$t$-channel stau propagator enhances the cross section at energies near the neutralino mass.
The effect of IB is seen in the dotted line of figure~\ref{fsusydif} where we plot 
the $df_{SUSY}/dE_{\gamma}$ as a function of  $E_{\gamma}$: with the  "bump"
at energies near the neutralino mass, $f_{SUSY}$ can be of the same order of magnitude as
the other models, in any case remaining 1--2 orders of magnitude smaller at lower energies.

We will use Model B as a reference in section~\ref{sec:dwarfsvsclusters} where we discuss the 
predictions for the total fluxes because it provides
the most optimistic scenario for detection.

\section{DM searches in dwarf galaxies} 	\label{sec:dwarfs}

\subsection{Selection of the sample}
Dwarf spheroidal (dSphs) satellites of the Milky Way are very good candidates for DM searches, as they are the most DM-dominated systems known in the Universe (with inferred mass-to-light (M/L) ratios as high as 1000 for some of the recently discovered dSphs) and are relatively close to us (less than 100 kpc in the case of Draco, UMi, and some of the new SDSS dwarfs). Moreover, most of them are expected to be free from bright astrophysical gamma sources, in contrast with other targets like the Galactic Center, nearby spiral galaxies, or very massive galaxy clusters.

Nowadays, we know the existence of at least 23 Milky Way satellites, more than half of them discovered 
in the last few years using SDSS data \cite{strigarinature}.  
Most of them have typically higher M/L ratios and are closer than the previously well-known dwarfs. 
Both facts should increase the chance of DM detection. Nevertheless,  observational data are still 
very scarce for most of them, so these numbers should be treated cautiously. A good example is Segue~1, 
which was catalogued as a unusually extended globular cluster soon after its discovery \cite{belokurov}, 
later as a new ultra-faint dwarf galaxy \cite{geha09,martinez09}, then again as a globular cluster 
\cite{niederste}, and recently new studies were published that favors the scenario of a highly 
DM-dominated dwarf galaxy \cite{martinez11,simon11}. %This last study applies a careful analysis of multi-epoch  velocity measurements by means of a Bayesian method that incorporates uncertainties due to imperfect  knowledge of membership, as well as the binary orbital motion of stars. The controversy, however, is probably still open and certainly even more spectroscopic data will be of crucial importance in clarifying this issue. 
The same might still happen to other similar objects such as Willman~1. %, for which it is also  evident that more observational data are needed in order to definitely dispel any doubts about its real nature. 
At present, the latest word on Willman~1 points towards a dwarf galaxy, probably with  
 irregular kinematics \cite{willmannew}. However, the main concern regarding Willman~1 is actually 
 linked to the more fundamental question of whether this object is gravitationally bound or not. If 
 Willman~1 is being disrupted by the Milky Way, the velocities are not a measurement of the object's 
 mass. This is in contrast with Segue~1, where no evidence of disruption is observed.  Thus the 
 velocities are tracing its gravitational potential, and therefore mass modeling (though uncertain) is valid. 

\begin{table}
\centering
\small
\begin{tabular}{llllccc}
  \hline
  \hline
DSph & D$_{\odot}$ (kpc) & L ($10^{3}~L_{\odot}$) & M/L ratio & Reference & Best IACTs & Observed? \\
\hline
Segue 1 & 23 & 0.3 & $>$1000 & \cite{geha09} & M, V & M \\
Sagittarius & 24 & 58000 & 25 & \cite{mateo98,helmiwhite01}  & H & H\\
UMa II & 32 & 2.8 & 1100 & \cite{simongeha07} & M, V & -\\
Willman 1 & 38 & 0.9 & 700  & \cite{simongeha07} & M, V & M,V \\
Coma Berenices & 44 & 2.6 & 450 & \cite{simongeha07} & M, V & -\\
UMi & 66 & 290 & 580 & \cite{mateo98} & M, V & W,V\\
Sculptor & 79 & 2200 & 7 & \cite{mateo98}  & H & H \\
Draco & 82 & 260 & 320 & \cite{mateo98}  & M,V & W,M,V \\
Sextans & 86 & 500 & 90 & \cite{mateo98}  & H & - \\
Carina & 101 & 430 & 40 & \cite{mateo98} & H & H \\
Fornax & 138 & 15500 & 10 & \cite{mateo98}  & H & - \\
  \hline
  \hline
\end{tabular}
\caption{\label{tab:dwarfs}A list of dSph satellites of the Milky Way that may represent the best candidates for DM searches according to their distance and/or inferred M/L ratio (the latter rather uncertain particularly for Segue~1, UMa~II and Willman~1). Dwarfs appear listed according to their distance (nearest first). Note that we included Sagittarius in the list even when it is not actually a dSph, given its traditional interest for DM searches (mainly due to its proximity). Columns (6) and (7) list, respectively, the best positioned IACT for observation and the IACT that already observed the object (in chronological order when more than one). Here, H = HESS; M = MAGIC; V = VERITAS; W = Whipple. }
\end{table}
\normalsize

We show in table \ref{tab:dwarfs} a tentative list of those dSph galaxies that could be the best candidates 
for DM searches at present according to their distance and/or inferred M/L ratio.  
Some of them have already been observed by IACTs with no success up to now 
\cite{MAGICdraco,SagHESS,whippleobs,MAGICwillman1,veritasdwarfs,hess09CMa,MAGICsegue,HESScarinasculptor}. 
The same negative results were recently obtained from Fermi data at lower energies \cite{FermiDM,Fermi_lines}.
Nevertheless, the search of the best target should not only be based on the expected level of the DM annihilation flux. 
Indeed, there are other aspects that might be crucial, such as instrumental effects (e.g., the telescope PSF), expected backgrounds in the 
direction of the object, uncertainties in the determination of the DM density profile, angular extension 
of the expected gamma signal, etc. We perform a detailed study in the next subsection, where only four out of 
all the dwarfs listed in table \ref{tab:dwarfs} were selected, mainly according to their distances and 
mass-to-light ratios. We chose two ``classical'' dwarfs, namely Draco and Ursa Minor, 
and two dwarfs recently discovered with SDSS data, i.e., Willman~1 and Segue~1. We took into account 
previous estimates of the DM annihilation flux in order to improve the selection 
\cite{Pieri08,strigari,masc,doro09,martinez09,martinez11,essig10}. The same study might be performed for the rest of dwarfs 
as well, but we note that the main conclusions will not probably change importantly (e.g.~\cite{Pieri08}).

\subsection{Looking for the best candidate}

The first step is to compute the gamma-ray DM annihilation flux profiles, as given by the astrophysical 
factor defined in section \ref{sec:flux}. %The flux profiles represent a very useful tool for quickly understanding how the expected DM annihilation flux is distributed in the object and which are the best candidate, at least to first order. 
To do so, we first need to model the DM distribution. Table \ref{tab:dSphs_param} summarizes the parameters that describe the DM density profiles of the four dwarfs under study. In the following we give more details on the adopted halo models.

\begin{table}[!h]
\centering
  \begin{tabular}{lccccccccccc}
  \hline
  \hline
DSph & Profile & r$_s$ (kpc) & r$_t$ (kpc) & $\rho_0$  (M$_\odot$~kpc$^{-3}$) & Ref.  \\
\hline
Draco-cusp & Kazantzidis & 1.189 & 1.6 & 3.1$\times$10$^7$  & \cite{masc} \\
Draco-core & Kazantzidis & 0.238& 1.6 & 3.6$\times$10$^8$ & \cite{masc} \\
UMi-A & NFW & 0.63 & 1.5 & 10$^8$ & \cite{StrigariPRD} \\
UMi-B & NFW & 3.1 & 1.5 & 10$^7$ & \cite{StrigariPRD} \\
Willman~1 & NFW & 0.173 & 0.9  & 4$\times$10$^8$ & \cite{strigari} \\
Segue~1 & Einasto & 0.07 & 0.8  & 10$^8$ & \cite{martinez09} \\
  \hline
  \hline
\end{tabular}
\caption{\label{tab:dSphs_param}Parameters that describe the DM density profiles of the four dwarf galaxies selected. [Note: In the draco-core case, $\rho_0$ is in units of M$_\odot$~kpc$^{-2}$.]}
\end{table}

\noindent\paragraph{Halo models; tidal and core radii.}
 In the case of Draco, we chose the two DM models presented in 
Ref.~\cite{masc}, characterized respectively by a cusp and a core in the very center of the dwarf. 
Both of them are of the form of a Kazantzidis DM density profile \cite{kmmds}, which is essentially an 
NFW plus an exponential cut-off at large radii to account for the loss of mass due to tidal disruptions 
in the outskirts of the dwarf. For UMi, we followed Ref.~\cite{StrigariPRD} and performed our calculations for 
the two NFW profiles given in that study. In the case of Willman~1, we extracted the main parameters from 
Ref.~\cite{strigari}. The tidal radius r$_t$, however, was directly derived from the Roche criterion \cite{binney&tremaine}:
\begin{equation}
r_t = \left (\frac{M_{dSph}}{3~M_{MW}(<R_{dSph})}\right)^{1/3} \times D_{dSph - GC}
\label{eq:tidal}
\end{equation}
where $M_{dSph}$ is here the mass of Willman~1 (fixed to 10$^7~M_{\odot}$), 
and $M_{MW}(<R_{dwarf})$ is the enclosed mass of the Milky Way out to the distance of the dwarf. 
For the latter, we assumed an NFW DM density profile for our galaxy with a virial mass of 
$M_{vir}=1.07\times10^{12}~M_{\odot}$ and a concentration of $c_{MW}=11$ following 
Ref.~\cite{pradaPRL}. Finally, $D_{dwarf - GC}$ is the distance from the dwarf to the Galactic 
Center. A similar procedure was followed for Segue~1 as well, for which we took an Einasto DM 
density profile with index $\alpha = 0.1$ (see eq.(\ref{eq:einasto})). We note here that we checked 
the robustness of our results by varying the tidal radius of both objects. We obtained that, though 
important for the computation of the flux coming from the outskirts, we got exactly the same 
flux level in the central regions, from which the maximum contributions to the flux are expected. 
Therefore, the uncertainties in the flux introduced by uncertainties in the computation of the tidal 
radius are in practice very small and completely negligible when assuming realistic values for it.

The integral of the square of the DM density along the line of sight at all angles ranging from 
zero up to the angle subtended by the tidal radius diverges at angles $\Psi_0 \rightarrow 0$ 
both for the NFW and the Kazantzidis profiles.
The usual approach is to assume a small constant 
DM core in the very center of the DM halo that prevents the divergence of the profile. 
The radius $r_{core}$ at which the self annihilation rate $t_l \sim (\langle\sigma_{ann}v\rangle 
n_{\chi}~r_{core})^{-1}$ equals the dynamical time of the halo $t_{dyn} \sim 
(G~\overline{\rho})^{-1/2}$, where $\overline{\rho}$ is the mean halo density and $n_{\chi}$ is the
neutralino number density, is usually taken as the radius of this constant density 
core~\cite{fornengo}. 
We used  $r_{core} =10^{-8}$~kpc in all our computations. We found that, in the range 10$^{-8}$~kpc~$< r_{core} < 0.1$~kpc, %and found that 
the total flux is almost insensitive to any value 
below $\sim$10$^{-3}$~kpc. Even when $r_{core}$ takes a value as high (and improbable for these 
objects) as 0.1 kpc, the total flux only changes by $\sim$10\%. These results are in concordance 
with those found in Ref.~\cite{fornengo} as well. The conclusion is that uncertainties coming from  
ignorance of the real size of the core radius will be likely negligible when assuming realistic 
values for it.

\begin{figure}[t!]
\centering
\includegraphics*[scale=0.6]{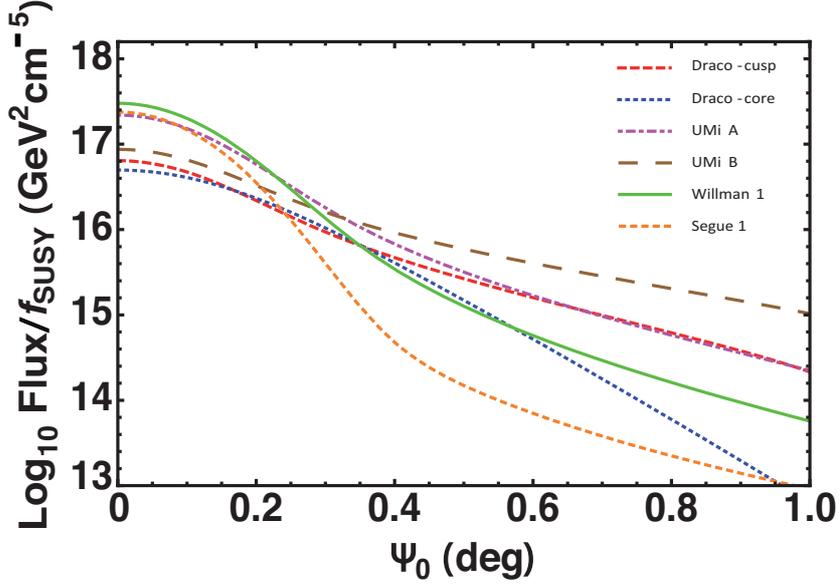}
\caption{ %\small
{Gamma-ray DM annihilation flux profiles, normalized to $f_{SUSY}$, for Draco, Ursa Minor, 
Willman~1, and Segue~1. The profiles were computed using those parameters listed in
 table \ref{tab:dSphs_param} for the DM distribution and assuming a PSF$=0.1^\circ$. From top to 
 bottom at $\Psi_0=0^\circ$, the profiles correspond to Willman~1, Segue~1, UMi-A, UMi-B, Draco-cusp, and 
 Draco-core following the nomenclature given in table \ref{tab:dSphs_param}.}} 
\label{fig:dwarfprofiles}
\end{figure}

\noindent\paragraph{Flux profiles and role of the PSF.}

The flux profiles of the four dwarfs are shown in figure \ref{fig:dwarfprofiles}. They were computed assuming PSF = $0.1^\circ$ and a DM distribution described by those parameters listed in table \ref{tab:dSphs_param} for each particular dwarf. From this figure, we can see that, according to the gamma-ray 
annihilation flux profiles, Willman~1 is the best candidate, since it reaches the highest annihilation
 fluxes at the center. Indeed, it reaches a factor of $\sim$1.3 more flux than the second best 
 candidate, Segue~1, and a factor 1.4 more than UMi-A. These three cases share therefore very similar 
 fluxes, while the others show a central flux which is more than a factor 3 lower in the best 
 case (UMi-B). We recall, however, that these results should be treated with caution, as the DM modeling of Willman~1 and Segue~1 may be subject to large uncertainties. %\cite{martinez09,MAGICwillman1,essig10}. %For  the former object, it is  not even clear whether the system is gravitationally bound or not, which might crucially affect its mass modeling. 

\begin{table}[!t]
\centering
\small
\begin{tabular}{lccccccccc}
  \hline
  \hline
DSph & Log$_{10}$ J$_T$ & $\psi_{90}$ & r$_{90}$/r$_s$ & J$_{01}$/J$_T$ & r$_{01}$/r$_s$ & $\psi_{r_s}$ & J$_{r_s}$/J$_T$ & Rank$_{01}$ & Rank$_{90}$   \\
 & (GeV$^2$cm$^{-5})$ & (deg) & & & & (deg) & & &\\
\hline
Draco-cusp & 17.57 & 0.65 &  0.77 & 0.149 & 0.12 & 0.85 &  0.96 & 5 & 5 \\
Draco-core & 17.48	  &   0.43 &  2.53 & 0.148	&  0.59 & 0.17 & 0.36 & 6 & 6 \\
UMi-A & 17.92 & 0.47 & 0.86 & 0.221 & 0.18 & 0.55 & 0.93 & 3 & 2 \\
UMi-B & 17.85 & 0.87 & 0.32 & 0.108 & 0.04 & 2.69 & 1.00 & 4 & 3 \\
Willman~1 & 17.93 & 0.31 & 1.18 & 0.292 & 0.38 & 0.26 & 0.84 & 1 & 1 \\
Segue~1 & 17.71 & 0.23 & 1.34 & 0.369 & 0.58 & 0.17 & 0.74 & 2 & 4 \\
  \hline
  \hline
\end{tabular}
\caption{\label{tab:dSphs_results}Value of the parameters that describe the characteristics of the DM-induced gamma emission in our sample of dwarf galaxies. See section \ref{subsub} for details on their definition and usefulness. This table was computed assuming a PSF$=0.1^{\circ}$.}
\end{table}
\normalsize

We further compute the parameters defined in section~\ref{subsub} and give them in table~\ref{tab:dSphs_results}. According to both Rank$_{01}$ and Rank$_{90}$ in this table, 
Willman~1 represents the best choice. Note that Segue~1 arises as the second best candidate according to its J$_{01}$, while it represents only the fourth best choice according to Rank$_{90}$. This is mainly due to the interplay between 
the DM distribution inside the objects, their distances, and especially the PSF of the IACT, which 
modifies the compactness of the signal emission and in this case leads to this interesting result. 

 Actually, the PSF plays a crucial role in correctly interpreting a possible DM annihilation signal. 
For instance, in Ref.~\cite{masc} the 
authors showed that it might be impossible to differentiate between a cuspy and a cored DM density 
profile in the case of having an insufficiently good PSF for Draco. Therefore, since the PSF 
drastically modifies the observed flux profiles, understanding  how the flux is {\it apparently} 
distributed in the object for a given PSF value becomes a crucial task. 
For example,  we mentioned above that, assuming an 
NFW profile, typically 90\% of the flux comes from the region inside r$_s$ (see table~\ref{tab:dSphs_results}). However, this is only strictly valid for ideal instruments 
where the PSF is ideally small. In practice, the larger the value of the PSF the larger the size of 
this region. 
Similarly, the observed annihilation flux coming from the inner 0.1$^{\circ}$ will decrease when 
worsening the PSF. To illustrate this point, we show in table~\ref{tab:psfs} the result of assuming 
different PSFs on the values of J$_{r_s}$, $\psi_{90}$ and the flux from the inner 0.1$\grado$. We chose the Draco-cusp case for this example.

\begin{table}
\centering
  \vspace{0.2cm}
  \begin{tabular}{lccc}
  \hline
  \hline
PSF (deg) & J$_{r_s}$/J$_T$ & $\psi_{90}$ (deg) & Flux($<0.1\grado$)/ Total flux  \\
\hline
0.01 & 0.97 & 0.63 & 0.29 \\
0.1 & 0.96 & 0.66 &  0.15\\
0.3 & 0.89 & 0.86 &  0.04 \\
1 & 0.29 & 1.01 & 0.01\\
  \hline
  \hline
\end{tabular}
\caption{\label{tab:psfs} Effect of the PSF on the {\it apparent} distribution of the annihilation flux as may be observed in the Draco dwarf, using the cuspy DM density profile of table~\ref{tab:dSphs_param}. The larger the value of the PSF, the larger the size of the region where 90\% of the flux is {\it apparently} emitted and the smaller the observed flux in the inner 0.1$^{\circ}$. See text for details.}
\end{table}

Another important fact that can be read from table~\ref{tab:dSphs_results} is that 
dwarf galaxies are not point-like according to the morphology of their annihilation 
signal. In particular, we find that J$_{01}$/J$_T$ only exceeds 30\% in the case of Segue~1, 
meaning rather extended emissions in all cases. Therefore, the typical assumption of treating dwarf 
galaxies as point-like sources seems  not to be very well founded and this fact should be ideally 
taken into account when analysing IACT data.

\begin{figure}
%\centering
%  \begin{minipage}[b]{0.49\textwidth}
%    \centering
\includegraphics[scale=0.4]{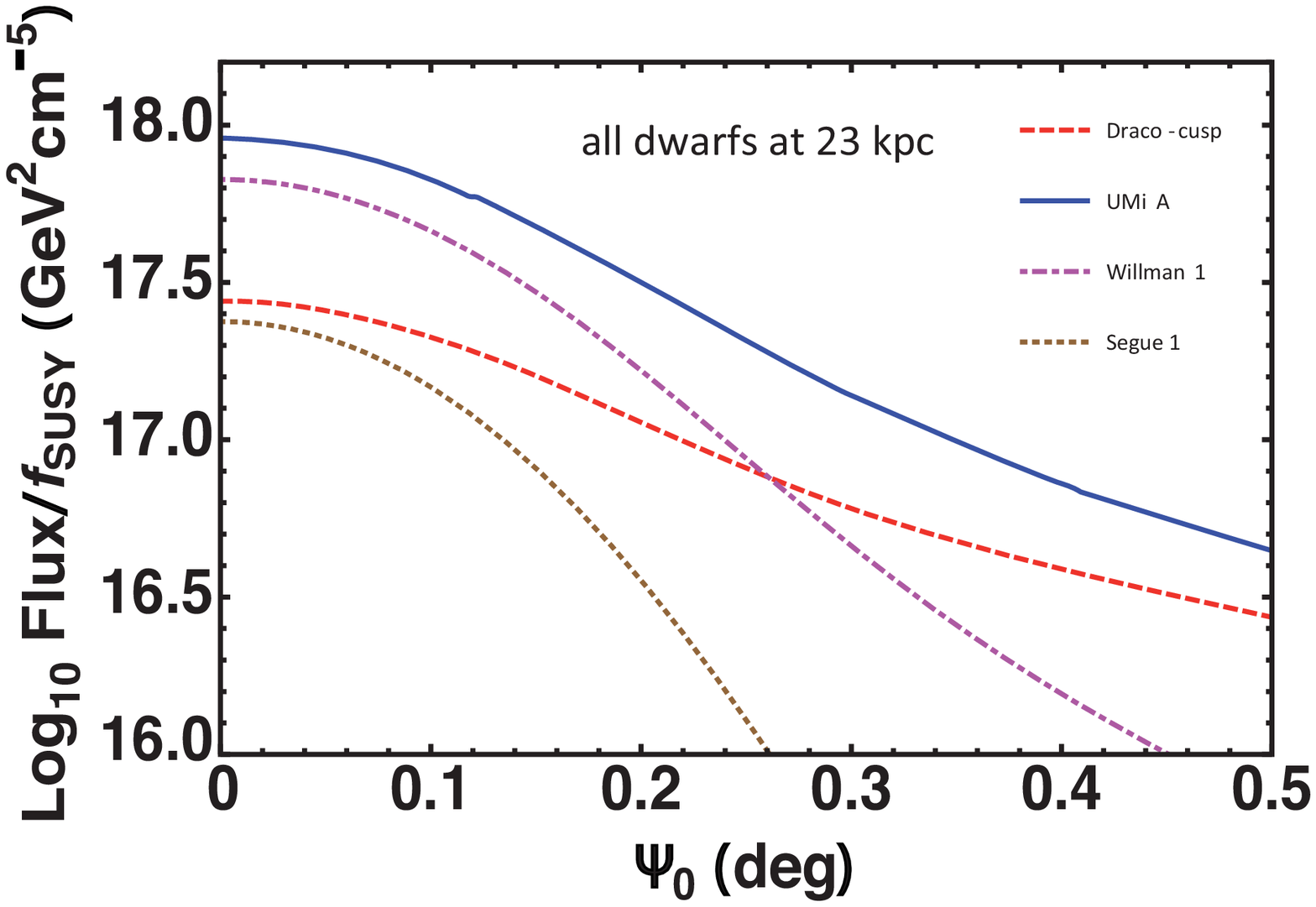}
%  \end{minipage}\\
%  \begin{minipage}[b]{0.49\textwidth}
%    \centering
\includegraphics[scale=0.4]{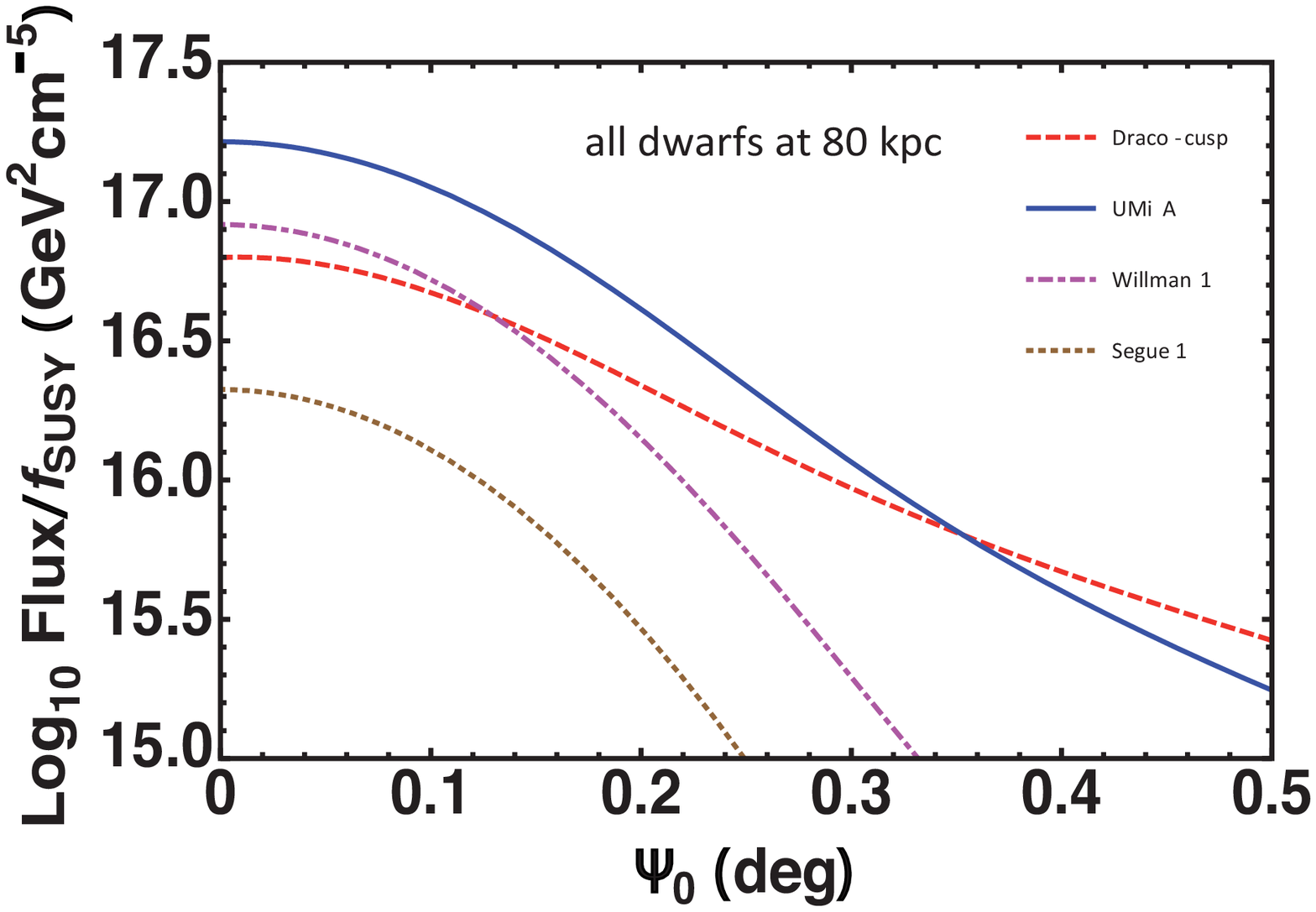}
%  \end{minipage}
\caption{{\it Left panel:} DM annihilation flux profiles (normalized to f$_{SUSY}$) for Draco-cusp, UMi-A, Willman~1 and Segue~1 in the case of placing all of them at a distance of 23 kpc (Segue~1 distance). {\it Right panel:} All dwarfs placed at 80 kpc (Draco distance). Differences in flux, although relevant, do not exceed a factor of a few both for the maximum annihilation flux reached at $\Psi_0=0^{\circ}$ and for the value of J$_{01}$; see text for discussion.} %}
\label{fig:dwarfs_distance}
\end{figure}

A final question to be addressed here refers to how important are the uncertainties introduced in the computation of the DM annihilation flux due to an incomplete knowledge of the real DM density profile. Implicitly, this was already done by assuming different DM density profiles for the same object (Draco and UMi). From table~\ref{tab:dSphs_results} and figure~\ref{fig:dwarfprofiles}, we conclude that the maximum of the annihilation flux (i.e., its value at $\Psi_0=0^{\circ}$) does not vary by more than a factor 1.3 between the cuspy and cored DM density profiles considered for Draco,
 and a factor 2.5 for the UMi-A and UMi-B models, respectively. As for J$_{01}$, probably the most relevant
  parameter, the ratios are basically the same, i.e., 1.2 and 2.4 respectively. There are also other 
 ways to quantify this issue. In figure~\ref{fig:dwarfs_distance} we performed the 
  exercise of placing Draco (using the cuspy profile), UMi (model UMi-A), Willman~1 and Segue~1 at the 
  same distance. We chose two distances: 23 kpc (Segue~1 real distance) and repeated the exercise for 
  80 kpc as well (Draco distance). Note that, though with different masses and different DM density profiles, all these dwarfs intriguingly show roughly comparable fluxes. More precisely, the maximum differences in flux (i.e., between UMi-A and Segue~1) at $\Psi_0=0^{\circ}$ are a factor $\sim$4 
  and $\sim$8 in the case that they are located at 23 and 80 kpc respectively. The situation is rather similar 
  when dealing with J$_{01}$, for which we obtain maximum differences of $\sim$6 and $\sim$9 in the case of 
  placing UMi-A and Segue~1 at 23 and 80 kpc, respectively. %Yet, these differences are not as great as one  may think. Indeed, in practice we will be dominated by a different kind of uncertainty: that coming from the fact that we do not know the exact DM particle physics model that corresponds to reality, which as shown in section \ref{sec:fsusy} can span over several orders of magnitude in $\langle \sigma v \rangle $. On the other hand,
The reason for these relatively small uncertainties introduced by the lack of knowledge of the exact DM density profile is probably that all these objects are dSph galaxies and therefore all share roughly similar physical properties. This is true at least for those parameters which are more     relevant for the computation of the DM annihilation flux, namely $\rho_0$ and r$_s$, which depend directly on the amount and distribution of matter in the innermost regions of these galaxies \cite{strigarinature}. 

\section{DM searches in galaxy clusters} 	
\label{sec:clusters}

\subsection{The selection of our sample}

In the currently accepted $\Lambda$CDM cosmological model, large-scale structures grow 
hierarchically through the merging of smaller systems into larger ones (e.g. \cite{peebles}). With 
masses of the order of 10$^{14}$-10$^{15}$ $\msun$ and radii of few Mpc, clusters of galaxies 
are the latest and most massive gravitationally bound objects to form in the Universe. Their 
mass is constituted principally of galaxies, gas and DM for roughly 5, 15, and 80\%, respectively 
(see Ref.~\cite{voit05} for a review). %Although other DM candidates, such as the dSph galaxy satellites of the Milky Way, are closer to us, the large amount of DM present in clusters makes them very attractive targets for DM searches as well. 
No clusters have yet 
been detected as gamma-ray sources, however they are expected to be significant gamma-ray emitters 
through conventional physical processes (see, for example, the review of Ref.~\cite{blasi} and also 
\cite{pinzke&pfrommer} for recent predictions). For DM purposes, it will be necessary to understand 
and to model this non-exotic emission in order to discriminate it from a possible DM annihilation signal. 

Some observations of clusters have been performed by the IACTs currently in operation, but only 
upper limits (ULs) have been obtained: from MAGIC on Perseus \cite{MAGICperseus}; from HESS on Coma 
\cite{hesscoma1,hesscoma2}, Abell 496, and Abell 85 \cite{hessabell}; from Whipple on Perseus and 
Abell 2029 \cite{whippleclusters}; from VERITAS on Coma and Perseus \cite{VERITAScluster1, 
VERITAScluster2} and from CANGAROO on Abell 3667 and Abell 4038 \cite{cangaroocluster}. 
All of them were preceded by EGRET ULs at lower energies~\cite{egretlimits}. 
The Fermi collaboration has already set ULs to the photon flux in the range 0.2--100 GeV 
for a sample of 33 promising galaxy clusters using 18 months of Fermi/LAT data \cite{Fermiclusters}, 
and it has analyzed some of them in the context of DM searches \cite{FermiDMinclusters}.

Table \ref{tab:clusters_param} shows our sample of galaxy clusters and their physical 
parameters. These clusters were selected mainly according to their distance and mass 
but also taking into account a good knowledge of those parameters needed to compute the 
DM density profiles, which were modeled here using the NFW profile in all cases. Moreover, 
we followed \cite{jeltema&profumo} in order to select the galaxy clusters that may be most
 interesting for gamma-ray DM searches. In that study, the authors calculated the DM-to-cosmic ray
  gamma-ray emission (see section \ref{sec:cosmicrays}) for the 110 clusters contained in the
   extended HIGFLUCS catalog \cite{RB02}. Their conclusion is that NGC5846 and NGC5813 reach the
    highest ratios, therefore probably representing  the best galaxy clusters to look for DM. Both 
    clusters are included in our sample for this reason. In addition, they identified Fornax as the 
    best target according to its expected DM annihilation flux only, and so we also included it
     in our study. Finally, we also picked Perseus, Coma, and Ophiuchus, even though they are not so 
     well considered according to these rankings. However, these galaxy clusters, plus Virgo, 
    have  traditionally captured a great deal of attention both in observational campaigns and theoretical 
     studies of diverse natures, given their proximity and masses. Therefore, they were selected for our galaxy cluster sample as well.

\begin{table}[t!]
\resizebox{\textwidth}{!}{
\begin{tabular}{cccccccccc}
  \hline
  \hline
Cluster &  z & D$_L$ & M$_{200}$ & R$_{200}$ & c$_{200}$ & r$_s$ & $\rho_0$ & Best IACTs & Observed? \\
(1) & (2) & (3) & (4) & (5) & (6) & (7) & (8) & (9) & (10)  \\
\hline
Perseus & 0.0183 & 77.7 & 7.71 & 1.90 & 3.99 & 0.477 & 7.25 & M,V & W,V,M   \\
Coma &  0.0232 & 101.2 & 13.84 & 2.30 & 3.78 & 0.609 & 6.44 & M,V & H,V \\
Ophiuchus & 0.0280 & 122.6 & 23.16 & 2.74 & 3.60 & 0.760 & 5.81 & H &  -\\
Virgo &  0.0036 & 15.4 & 5.6 & 1.68 & 4.21 & 0.433 & 6.81 & M,V,H & -\\
Fornax &  0.0046 & 19.8 & 1.01 & 0.96 & 4.80 & 0.201 & 11.0 & H & - \\
NGC5813 & 0.0064 & 27.5 & 0.27 & 0.62 & 5.42 & 0.115 & 14.5 & M,V,H & - \\
NGC5846 & 0.0061 & 26.3 & 0.38 & 0.69 & 5.26 & 0.132 & 13.5 & M,V,H & -\\
\hline
\hline
\end{tabular}}
\caption{\label{tab:clusters_param}Main physical parameters of the galaxy clusters selected for this study.
\emph{Columns}: (1) Cluster name; (2) redshift; (3) Luminous distance in Mpc; (4) M$_{200}$ in units 
of 10$^{14} M_\odot$; (5) R$_{200}$ in Mpc; (6) c$_{200}$ calculated assuming the c$_{200}$(M$_{200}$) 
relation given in Ref.\cite{duffy08} for relaxed DM halos; (7) scale radius in Mpc; (8) $\rho_0$ 
in units of 10$^{14}$ M$_\odot$~Mpc$^{-3}$. An NFW DM density profile was assumed to derive r$_s$ and 
$\rho_0$; (9) Best positioned IACT for observation: H = HESS; M = MAGIC; V = VERITAS; W = Whipple; (10) IACT that 
already observed the object (in chronological order when more than one). \emph{ References}: (2) and (4)-(5) from Ref.~\cite{RB02} for all the clusters but Virgo, for 
which we used the M$_{102}$ and R$_{102}$ values given in Ref.~\cite{fouque01}, that were later converted to the M$_{200}$ and R$_{200}$ given above.}%; (3) computed using the online application given in htttp://www.astro.ucla.edu/\textasciitilde wright/CosmoCalc.html.}
\end{table}
 
\subsection{Looking for the best candidate}

\begin{figure}
\centering
\includegraphics*[scale=0.6]{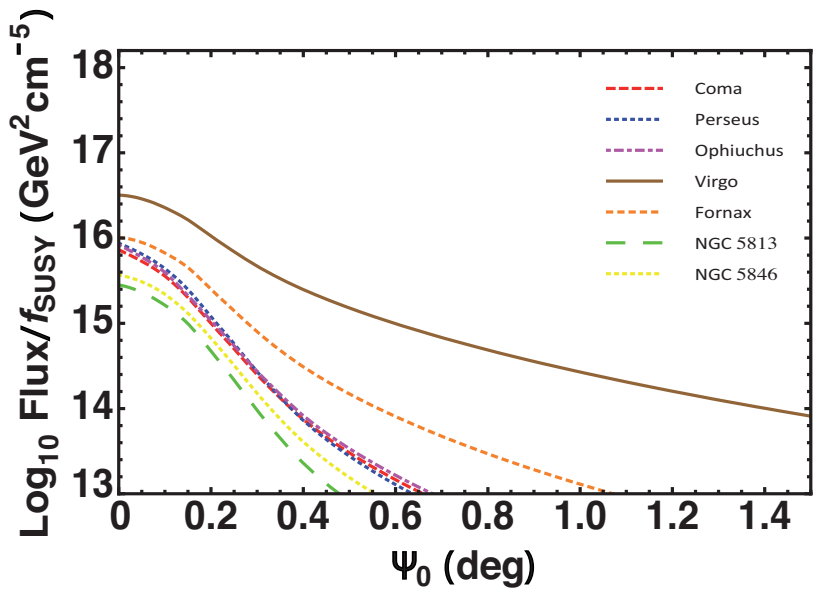}
\caption{Gamma-ray DM annihilation flux profiles, normalized to $f_{SUSY}$, for Virgo, Fornax, Perseus, Ophiuchus, Coma, NGC5846 and NGC5813 (from top to bottom at $\Psi_0=0^\circ$). The profiles were computed using those parameters listed in table \ref{tab:clusters_param} for the DM density profiles and assuming a PSF $=0.1^\circ$.}  
\label{fig:clustersprofiles}
\end{figure}

Figure~\ref{fig:clustersprofiles} shows the flux profiles for the sample of galaxy clusters 
calculated using the parameters listed in table~\ref{tab:clusters_param}.
Virgo is  the galaxy cluster that yields by far the highest DM 
annihilation fluxes at all angles $\Psi_0$ in our sample. Fornax is the second best candidate, 
while Perseus, Ophiuchus and Coma, which show very similar DM annihilation flux profiles despite 
their different distances, are below Fornax by a factor of a few in the central regions. Finally, 
the two NGC clusters in the sample are the  clusters that yield the lowest fluxes, these 
being already more than an order of magnitude below Virgo at $\Psi_0=0\grado$. 

In table~\ref{tab:clusters_results} we report the values of the benchmark quantities defined in 
section \ref{subsub}. There are no surprises with respect to 
figure~\ref{fig:clustersprofiles} regarding the best and worst candidates in the sample (see 
Rank$_{01}$ and Rank$_{90}$). Probably, the main conclusion that can be extracted from this table 
is that we should expect the induced gamma-ray DM emission in these objects to be quite 
extended for IACTs, as $\psi_{90}$ (or alternatively $\psi_{r_s}$, which contains in most cases 
$\sim$90\% of the flux) is typically larger than 0.3$^\circ$, i.e., three 
times the usual PSF of these instruments. Indeed, the inner 0.1$^\circ$ rarely encloses more than 
$\sim$30\% of the total DM annihilation flux (see J$_{01}$/J$_T$). We note that rather similar 
results were achieved for dwarfs as well (see table \ref{tab:dSphs_results}).

\begin{table}[t!]
\resizebox{\textwidth}{!}{
  \begin{tabular}{lccccccccc}
  \hline
  \hline
Cluster & Log$_{10}$ J$_T$ & $\psi_{90}$ & r$_{90}$/r$_s$ & J$_{01}$/J$_T$ & r$_{01}$/r$_s$ & $\psi_{r_s}$ & J$_{r_s}$/J$_T$ & Rank$_{01}$ & Rank$_{90}$   \\
& (GeV$^2$cm$^{-5})$ & (deg) & & & & (deg) & & &\\
\hline
Perseus & 16.25 & 0.31 &  1.08 & 0.323 & 0.35 & 0.29 &  0.88 & 3 & 3 \\
Coma & 16.18	  &   0.34 &  0.99 & 0.312	&  0.29 & 0.34 & 0.90 & 5 & 5 \\
Ophiuchus & 16.22 & 0.35 & 0.98 & 0.318 & 0.28 & 0.36 & 0.91 & 4 & 4 \\
Virgo & 17.37 & 1.45 & 0.90 & 0.113 & 0.06 & 1.61 & 0.92 & 1 & 1 \\
Fornax & 16.58 & 0.55 & 0.94 & 0.214 & 0.17 & 0.58 & 0.91 & 2 & 2 \\
NGC5813 & 15.82 & 0.29 & 1.22 & 0.313 & 0.42 & 0.24 & 0.83 & 7 & 7 \\
NGC5846 & 15.97 & 0.32 & 1.12 & 0.297 & 0.35 & 0.29 & 0.87 & 6 & 6 \\
  \hline
  \hline
\end{tabular}}
\caption{Value of the parameters that describe the characteristics of the DM-induced gamma 
emission in our sample of galaxy clusters. See section \ref{subsub} for details 
on their definition and usefulness. This table was computed assuming a PSF$=0.1^{\circ}$.}
\label{tab:clusters_results} 
\end{table}

\subsection{The effect of substructure} \label{sec:sub}

In the $\Lambda$CDM paradigm, the smallest dense halos form first and later 
merge to originate larger structures. This hierarchical scenario has as a direct
 consequence the presence of a large amount of substructure in CDM halos. As the 
 DM annihilation signal is proportional to the DM density squared, this clumpy 
 distribution of sub-halos inside larger halos may boost the DM annihilation flux 
 considerably. The flux enhancement will be more important for the most massive 
 halos as they enclose more hierarchical levels of the structure formation. Therefore, 
 it becomes essential to  quantify precisely the substructure boost when computing 
 the DM annihilation flux from galaxy clusters. In contrast, the effect will turn 
 out to be insignificant for dwarfs. 

The effect of substructures on the DM annihilation flux has already been studied 
both analytically (e.g., in Refs.~\cite{pieri08b,martinez09,afshordi}), and making 
use of state-of-the-art N-body cosmological simulations \cite{kuhlen08,springel}, 
although the exact calculation of the substructure boost has been challenging. It becomes 
quite difficult to  calculate analytically the survival probabilities of substructures
 within the host halos, while the most powerful N-body simulations fail to simulate 
 the sub-halo hierarchy below a mass $\sim$10$^5 \msun$, still very far from the 
 minimum halo mass predicted to be present in the structure formation scenario, of 
 the order of 10$^{-6}\msun$ or even smaller \cite{minmass}.

Recently, Kamionkowski, Koushiappas, and Kuhlen developed in Ref.~\cite{kkk} a 
semi-analytical model in order to  include the substructure in the computation 
of the DM annihilation flux (hereafter 3K10 model). Their model is based on a previous 
analytical study that described the self-similar substructure expected from hierarchical 
clustering \cite{kamion1}. The 3K10 model makes an upgrade of this first work by 
performing a calibration to the Via Lactea II N-body cosmological simulation \cite{VLII}. 
After this calibration it is possible to use the model to obtain a suitable extrapolation of the 
results of the simulation below its lower mass limit. In addition, the 3K10 model includes 
a good description of the distribution of substructure in the halo when varying the 
galactocentric radius. Both achievements make possible a more realistic and precise 
computation of the substructure boost to the DM annihilation flux. The 3K10 model indicates 
that the lower mass halos will not contribute greatly to this boost. 

In the 3K10 framework, the boost factor $B(r)$ is given by:
\begin{equation}
      B(r) = f_s e^{\Delta^2}  + 
      (1-f_s)\frac{1+\alpha}{1-\alpha} \left[
      \left(\frac{\rho_{max}}{\rho(r)} \right)^{1-\alpha}
      -1 \right].
\label{eqn:diffboost}
\end{equation}
where $f_s$ refers to the volume of the halo that is filled with a smooth dark matter 
component with density $\rho$(r), while the fraction $(1-f_s)$ corresponds to a high-density 
clumped component due to the presence of substructures. We chose $\Delta=0.2$, $\alpha=0$ 
and $\rho_{max}=80$ GeV~cm$^{-3}$, which are the  values found when calibrating the 3K10 
model to the VL-II simulation. We refer the reader to Ref.~\cite{kkk} for a detailed 
description of each of these terms. Note that the boost factor is indeed composed of 
two terms: a first term $B_s= f_s e^{\Delta^2}$ due to the finite width of the 
smooth component (that will have little importance here) and a second term due to substructures. 

We can also numerically evaluate the total boost from substructure within a radius $R$:
\begin{equation}
B(<\!R) = \frac{\int_0^R B(r) \, \rho^2(r) \, r^2 \, dr}{\int_0^R \rho^2 (r) \, r^2 \, dr}.
\label{eq:totalboost}
\end{equation}

By applying the 3K10 methodology, we assume to be independent	 of the host halo 
mass and the properties of the sub-halo population. The only uncertainty would come from 
the parameters of the model,  
especially from $f_s$, which is related to how effective tidal stripping (or any other sub-halo destroying 
mechanism) is. In the 3K10 formalism, $f_s$(r) was determined from VL-II, but the 
simulations are many orders of magnitude away from resolving the whole sub-halo hierarchy, 
and therefore $f_s$ is not known with unlimited precision. 

On the other hand, our intention 
is to apply the 3K10 formalism to dwarf galaxies and galaxy clusters and not only to MW-sized 
objects, so it is necessary to rescale $f_s$(r) in order to correctly accommodate 
it to halos of different sizes. We do so by replacing the $\rho(r=100\,{\rm kpc})$ parameter 
in eq.(4) of Ref.~\cite{kkk} by $\rho(r=3.56\times r_s~\,{\rm kpc})$, i.e.:
\begin{equation}
     1-f_s(r) = 7\times10^{-3} \left(
     \frac{\rho(r)}{\rho(r=3.56\times r_s~\,{\rm kpc})} \right)^{-0.26},
\label{eq:fs_mod}
\end{equation}
\noindent as $3.56$ is the ratio between the VL-II scale radius ($r_s=28.1$ kpc) 
and $r=100$ kpc (value extracted {\it ad hoc} from VL-II to properly calibrate the 
3K10 model). Note that in doing so we are assuming the same radial dependence of $f_s$ 
for all halo masses, only rescaling it to the particular size of the new object. 

\begin{figure}
\centering	
\includegraphics*[scale=0.6]{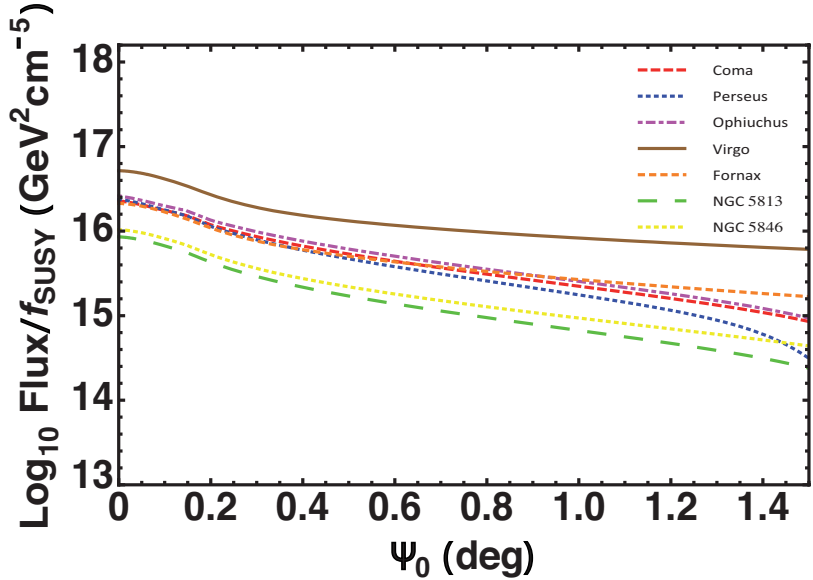}
\caption{Gamma-ray DM annihilation flux profiles, normalized to $f_{SUSY}$, for Perseus, 
Coma, Ophiuchus, Virgo, Fornax, NGC5813, and NGC5846. The profiles were computed using 
those parameters listed in table \ref{tab:clusters_param} for the DM density profiles and 
assuming a PSF=$0.1^\circ$. Substructure is  included here following the 3K10 model \cite{kkk} 
using those parameters given in the text. From top to down at $\Psi_0=1.2^\circ$, the profiles 
correspond to Virgo, Fornax, Ophiuchus, Coma, Perseus, NGC5846, and NGC5813.}  
\label{fig:clusters_sub}
\end{figure}

In figure \ref{fig:clusters_sub} we show the result of applying the 3K10 model to our 
sample of galaxy clusters using the values given above for $f_s$, $\rho_{max}$, and 
$\alpha$, as well as the new scaling relation introduced in eq.~(\ref{eq:fs_mod}). The 
substructure boost turns out to be extremely important in all cases, its effect being 
relevant at all l.o.s.\ angles $\Psi_0$. Note that the largest flux enhancements, however, are 
achieved at the largest $\Psi_0$ (compare with figure~\ref{fig:clustersprofiles}). Furthermore, Ophiuchus, Perseus, and Coma are now at the same flux level as Fornax. % thanks to the more prominent role of substructures in those clusters than in the latter.
The quantitative analysis is summarized in table~\ref{tab:clusters_results_sub}. 
$Rank_{01}$ and $Rank_{90}$ are now significantly
altered with respect to table~\ref{tab:clusters_results}.
The total boost within the virial radius
gives us an idea of the global importance of substructure for each object: typical values of 
this boost for the most massive halos in the sample are of the order of 50.

 Yet, there are important observational consequences that arise when comparing tables \ref{tab:clusters_results} and \ref{tab:clusters_results_sub}: 
J$_{r_s}$/J$_T$ rarely reaches values greater than 0.2 when including substructure, 
in contrast with the typical values $\sim$0.9 found without substructure. This means that the 
gamma-ray DM annihilation induced emission is indeed even less concentrated than previously 
thought, the object being significantly more extended for IACTs. We note that this fact already 
has important implications, e.g., on those conclusions achieved in Ref.~\cite{MAGICperseus} 
regarding DM searches in Perseus with the MAGIC telescope, where the authors assumed that the 
majority of the flux approximately comes from a region comparable with the telescope PSF and 
J$_{r_s}$/J$_T=0.9$. Other related quantities in table \ref{tab:clusters_results_sub} where 
this same issue is clearly visible are J$_{01}$/J$_T$, which surprisingly falls  below 
4\% for all the considered objects, and r$_{90}$/r$_s$, now of the order of 4--5 in contrast to 
the previous factor $\sim$1. Indeed, table \ref{tab:clusters_results_sub} shows that $\psi_{90}$ 
is, in all cases, somewhat greater than 1$^\circ$, clearly indicating the distinct extended 
nature of the gamma-ray emission. Similar conclusions have also been obtained in recent works adopting different substructure treatments \cite{pinzke11,gao11}. They found, however, much larger substructure boost factors (roughly a factor 20 higher) than those given in our table~\ref{tab:clusters_results_sub}.

\begin{table}[t!]
\centering
\resizebox{\textwidth}{!}{
  \begin{tabular}{lcccccccccc}
  \hline
  \hline
Cluster & B($<R_{vir}$) & Log$_{10}$ J$_T$  &  $\psi_{90}$  & r$_{90}$/r$_s$ & J$_{01}$/J$_T$ & r$_{01}$/r$_s$ & $\psi_{r_s}$ & J$_{r_s}$/J$_T$ & Rank$_{01}$ & Rank$_{90}$   \\
& & (GeV$^2$cm$^{-5})$ & (deg) & & & & (deg) & & &\\
\hline
Perseus & 34.0 & 17.73 & 1.22 &  4.24 & 0.037 & 0.14 & 0.29 &  0.19 & 3 & 5 \\
Coma & 51.6 & 17.84  &  1.41 &  4.08 & 0.028 & 0.29 & 0.34 & 0.20 & 4 & 4 \\
Ophiuchus & 54.0 & 17.89 & 1.38 & 3.89 & 0.028 & 0.28 & 0.36 & 0.21 & 2 & 3 \\
Virgo & 55.0 & 19.11 & 7.29 & 4.55 & 0.004 & 0.06 & 1.61 & 0.18 & 1 & 1 \\
Fornax & 39.9 & 18.17 & 2.97 & 5.11 & 0.013 & 0.17 & 0.58 & 0.16 & 5 & 2 \\
NGC5813 & 34.8 & 17.33 & 1.36 & 5.69 & 0.035 & 0.42 & 0.24 & 0.14 & 7 & 7 \\
NGC5846 & 36.1 & 17.51 & 1.59 & 5.54 & 0.028 & 0.35 & 0.29 & 0.15 & 6 & 6 \\
  \hline
  \hline
\end{tabular}}
\caption{Same as table~\ref{tab:clusters_results} but now including substructure. B($<R_{vir}$) 
is the total boost within the virial radius of the object, as given by eq.~(\ref{eq:totalboost}). 
This table was computed assuming a PSF$=0.1^{\circ}$.}
\label{tab:clusters_results_sub}
\end{table}
\normalsize

For completeness, we also studied the effect of substructure on our 
sample of dwarf galaxies, although, as mentioned, its importance is expected to be 
negligible for these objects. Effectively, we found the following values for the 
total boost, as given by eq.~(\ref{eq:totalboost}), within the tidal radius: 1.12, 1.12, 
1.16, 1.16, 1.19, 1.31 for Segue~1, Willman~1, UMi-A, UMi-B, Draco-cusp, and Draco-core 
respectively. The DM annihilation flux profiles are not significantly affected by introducing 
substructure either, except marginally in the outer regions, where in any case the level of 
the flux still remain extremely low.

\subsection{ $\gamma$-rays with a non-DM origin in clusters}   \label{sec:cosmicrays}
When considering DM searches in galaxy clusters one has to carefully consider the 
possible emission from other non-DM sources. In first place, some clusters contain 
bright active galactic nuclei (AGN) \cite{urry95} that may hinder the possible DM 
detection. These sources, while often detected at Fermi-LAT energies \cite{ngc1275}, are 
not always observed in the GeV-TeV range. This is due to the  high-energy emission cut-off 
given by the decreasing inverse Compton (IC) scattering efficiency in the Klein-Nishina regime. 
Moreover, in this sense,  the AGN jet inclination angle and the gamma-ray 
absorption in the source neighborhood also play an important role. However, many AGN are proved 
to  emit efficiently at very high energies and, additionally, they typically show variable 
emission. Therefore, no general conclusions on their impact on cluster DM searches can be drawn 
and their emission should be carefully modeled in order to correctly derive implications for DM 
(see Ref.~\cite{colafrancesco10}). Alternatively, AGN could be masked away as  
they are typically point-like objects for IACTs.

Another source of gamma-rays in clusters that can frustrate DM searches are cosmic rays 
(CR). CR electrons 
are directly visible in many galaxy clusters via synchrotron radiation in radio, forming 
the so-called cluster radio halos \cite{ferrari08}. These particles can be injected into 
the intra-cluster medium (ICM) by various sources such as structure formation shocks, 
radio galaxies and supernovae driven galactic winds (see the introduction of Ref.~\cite{MAGICperseus} 
and reference therein). The same sources can also inject CR protons into the ICM. Both 
CR protons and electrons can generate high-energy gamma-ray emission, via pion-decay 
\cite{volk96,ensslin97,pfrommer03,pfrommer04,pfrommer08,pfrommer08b} and IC up-scattering 
of the cosmic microwave background \cite{loeb00,totani00,miniati02,miniati03,petrosian08}, 
respectively. A detailed discussion of these particles and their relative contribution to 
the gamma-ray emission is beyond the scope of this study and we refer the reader to the above 
references as well as to Ref.~\cite{pinzke&pfrommer} and references therein.  Additionally, 
see also Ref.~\cite{brunetti10} for the effect of turbulence on merging clusters that 
can also accelerate CRs to very high-energies. 

State-of-the-art cluster cosmological simulations show that the dominant contribution to the 
CR induced gamma-ray emission, at the energies of interest here, should come from pion-decay 
\cite{pfrommer08,pinzke&pfrommer}. Spatially, this emission is concentrated in the inner part of the 
cluster as it is proportional to the squared ICM density. 
On the other hand, in this study we showed that 90\% of the DM emission in a cluster 
typically comes from within a radius of $\sim$0.3$^\circ$ (or larger) that dramatically 
changes to $\sim$1.2$^\circ$ (or larger) when including the substructure treatment (see
 tables 7 and 8). 
 
 One might think that the expected DM-induced emission spatial profile would 
 be comparable to the CR-induced one, but, in a realistic scenario that correctly includes 
 substructures, the former is clearly more extended and has a shallower profile (see for comparison 
 fig. 13 of Ref.~\cite{pinzke&pfrommer} and the recent \cite{pinzke11}). In practice, given that a typical Cherenkov telescope 
 PSF is $\sim$0.1$^\circ$, it is hard to imagine that existing IACTs will be able to distinguish 
 between CR and DM from the emission spatial profiles (this may change with the next-generation 
 Cherenkov Telescope Array). %Note that, in any case, these considerations could be critical for observational and data analysis strategies and deserve further study. 
 Fortunately, the 
 spectral profile of the CR-induced emission also helps. In the energy regime of interest for 
 IACTs, between 50 GeV and few TeV, this emission is expected to follow a power-law spectrum 
 with a spectral index of $-$2.2 \cite{pinzke&pfrommer} while the latter is expected to be harder ($-$1.5 
 or more) when having a DM origin. Therefore, should the DM annihilation emission be comparable 
 to the CR induced one, the very peculiar characteristics of the former would be hopefully 
 recognizable against the latter.

\section{DM annihilation flux predictions and detection prospects for IACTs} 	\label{sec:dwarfsvsclusters}

\subsection{Galaxy clusters or dwarf galaxies?} \label{sec:comparison}

\begin{figure}[t]
\centering
\includegraphics*[scale=0.43]{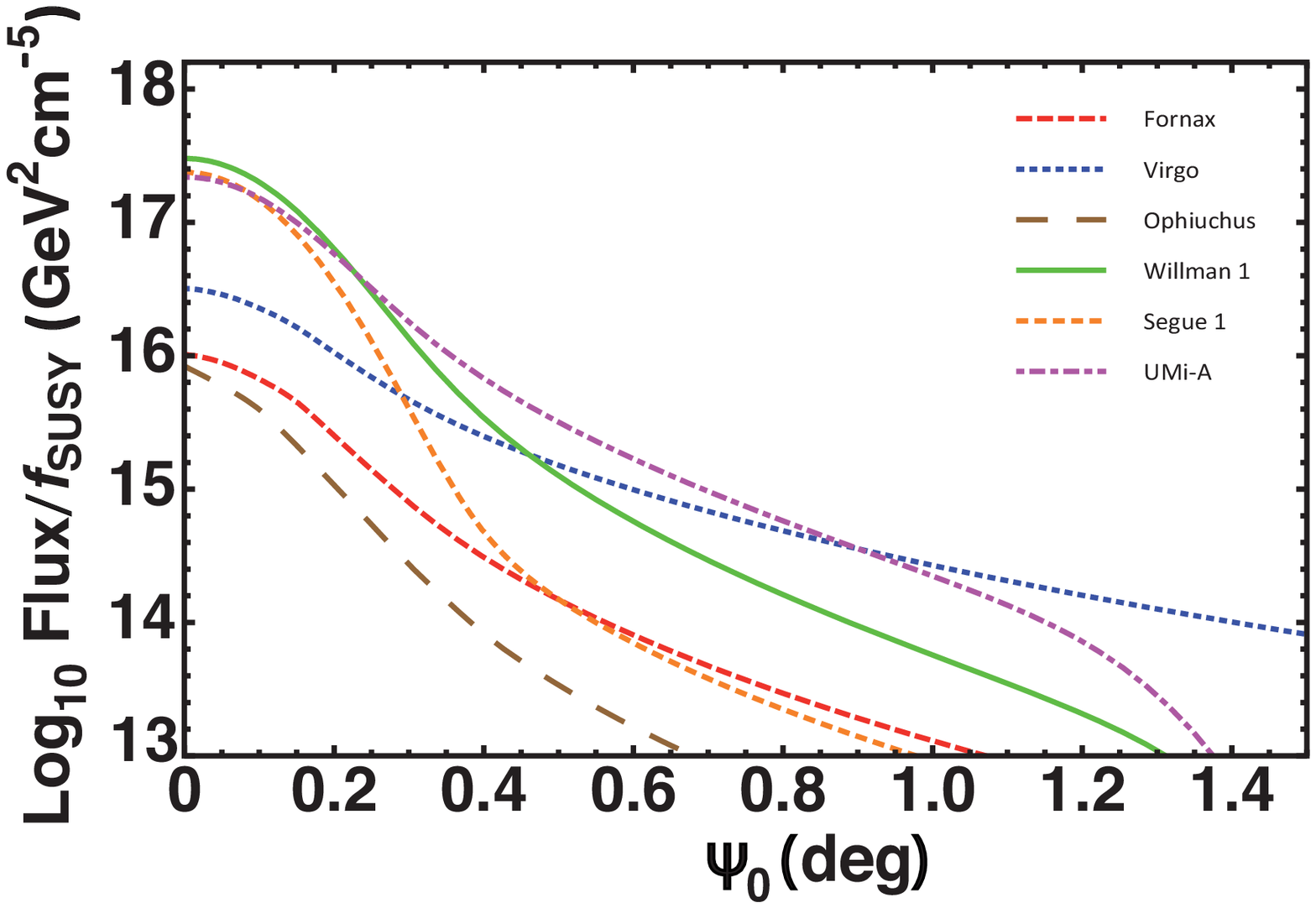}
\includegraphics*[scale=0.41]{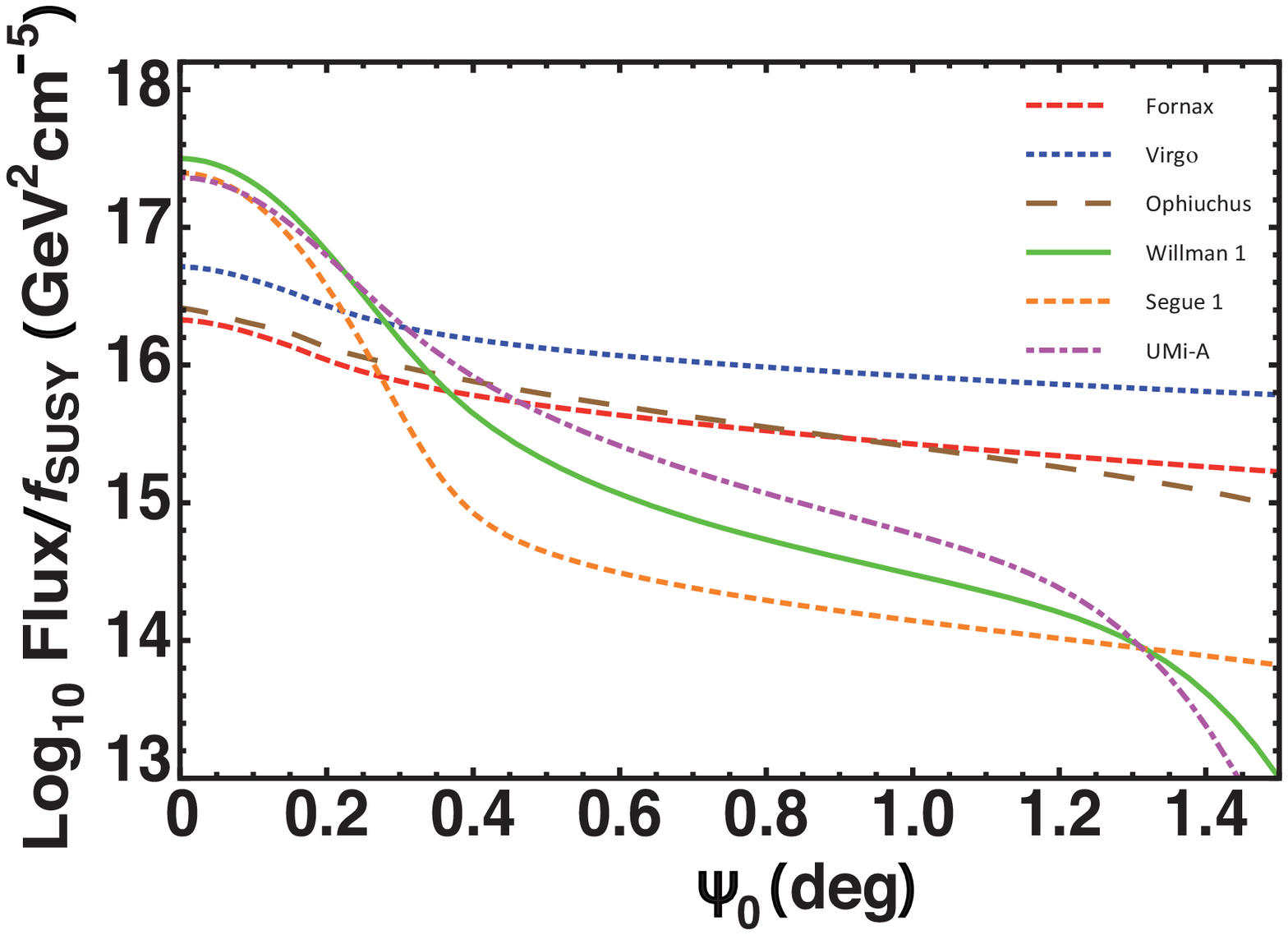}
\caption{{\it Left panel:} Comparison of the DM annihilation flux profiles (normalized 
to f$_{SUSY}$) for the subsample of those three dwarfs and three clusters with the 
highest fluxes. {\it Right panel:} Same as left panel but this time including substructure 
following the 3K10 model described in section \ref{sec:sub}.} %}
\label{fig:clustersVSdwarfs}
\end{figure}

In this section, we will compare the results previously obtained for dwarf galaxies 
with those obtained for galaxy clusters with the aim of elucidating 
the best candidates for gamma-ray DM searches.
The result of the comparison is given in figure~\ref{fig:clustersVSdwarfs}, where we show the case 
with no substructure at all (left panel) and a second case where we included substructure, 
in both dwarfs and clusters (right panel). For clarity, we do not use 
our whole sample of objects, but just the sub-sample composed by those three dwarfs 
---Willman~1, Segue~1 and UMi-A--- and 
three clusters ---Virgo, Fornax and Ophiuchus--- with the highest fluxes.

In both panels, dwarf galaxies reach the highest flux levels at $\Psi_0=0^\circ$, 
roughly an order of magnitude larger than those expected from clusters. This therefore
 seems to favor dwarfs against galaxy clusters, 
particularly for point-like based observational search strategies. However, note that galaxy clusters dominate the gamma-ray DM-induced emission at large angles 
once substructure is properly taken into account. This happens at radii greater than 
$\sim$0.4$^\circ$ in all cases, fluxes remaining substantially higher than those expected 
from dwarfs and decreasing quite slowly up to very large radii, contrary to what happens 
in dwarfs. Actually, once we include the effect of substructure, some of these galaxy clusters emit much 
more DM annihilation flux {\it in total} than the best dwarf galaxies. For example Virgo, as can be seen by comparing J$_T$ 
in tables~\ref{tab:dSphs_results} and~\ref{tab:clusters_results_sub}, 
gives a flux larger than Willman~1 by a factor $\sim$13. However, the main contribution to the total flux now comes from the outer regions, where the flux level is comparatively quite low with respect to that reached in the very center. Thus, if our search strategy can deal with quite extended sources (meaning $\sim1-1.5^\circ$, 
which, as shown in table \ref{tab:clusters_results_sub}, is the typical value of $\psi_{90}$, 
i.e., the typical size of the 90\% emitting region), then galaxy clusters probably are 
the best candidates or at least represent good competitors to dwarfs.

\subsection{J-values comparison with other works}
Below we comment on the agreement/disagreement of our J-values with those found in some works in the literature. We note that, when performing such a comparison, one has to be very careful in dealing with the different notations and definitions (see e.g. Appendix A in ref.~\cite{charbonnier11} for a useful discussion on conversion units and related issues).

\begin{itemize}
  \item {Dwarfs:}
  \begin{itemize}
     \item In the classical work of Ref.\cite{evans}, authors found a $J_{01}$ for Draco which is roughly a factor 1.5 higher than the one given in our table \ref{tab:dSphs_results} for the Draco-cusp case.
     \item After correcting by different definitions and angular apertures, we found ref.~\cite{strigari08} to predict a slightly lower $J_{02}$ value (i.e. eq.~(\ref{eq:integprof}) with $\psi=0.2\grado$) for Willman~1; more precisely they found 8\% less flux than the one we find. As for UMi-A, we obtain a slightly higher $J_{01}$ value. 
      \item We obtain similar $J_{01}$ values for UMi-A and UMi-B  than those given in Ref.~\cite{Pieri08}. However, we end up with significantly lower values for Draco. The found difference is completely attributable to the different halo parameters used in each case.
    \item Authors in ref.~\cite{essig10} find a $J_{01}$ value for Segue~1 which is a factor 1.6 larger than the one given in our table~\ref{tab:dSphs_param}.
    \item Our $J_T$ values for both Draco and Willman~1 are in good agreement with those given in ref.~\cite{doro09} for the same objects. More precisely, authors in that work find Draco and Willman~1 to yield a factor 1.1 and 1.3 more DM annihilation flux respectively.
    \item The MAGIC collaboration used a $J_{01}$ value for Segue~1 in ref.~\cite{MAGICsegue} that is a factor 2.8 higher than the one given in this work. These differences are due to the slightly different halo parameters used in each case.
    \item The results shown in our table~\ref{tab:dSphs_param} for both Draco-cusp and UMi-A match perfectly with those given in the recent ref.~\cite{charbonnier11} for the same objects.
     \end{itemize}
  \item{Clusters:} In case of clusters, only absolute flux values have been quoted in the scarce literature available so far, rather than providing J-values for them. Thus, a one-to-one comparison is not straightforward as one may take into account the particular particle physics model used in each case. Below, we perform just a qualitative comparison.
  \begin{itemize}
  	\item Ref.~\cite{jeltema&profumo} ranks the galaxy clusters according to their expected DM annihilation flux in their table VIII. Limiting here the comparison to those clusters that are common to both works, they found, starting from the brightest, the following: Fornax, Ophiuchus, Coma, Virgo, Perseus, NGC5846, NGC5813. In our work, we find (with substructure, table \ref{tab:clusters_results_sub}): Virgo, Fornax, Ophiuchus, Coma, Perseus, NGC5846, NGC5813. Therefore, except for Virgo, for which we find a higher flux, both works share a similar ranking list. On the other hand, note that, although authors in ref.~\cite{jeltema&profumo} quoted NGC5846 and NGC5813 as the best ones for DM searches given their comparatively low cosmic ray induced gamma-ray emission (see their Fig.16), our results are actually perfectly compatible. Attending {\it only} to their DM annihilation flux, both galaxy clusters are not by far the most promising ones for DM searches, neither in \cite{jeltema&profumo} nor in our work.\footnote{In any case, we found it convenient to include both NGCs in our cluster sample and discuss their flux predictions in the context of a joint study dwarfs+clusters, keeping in mind that, as we do not perform a treatment of the cosmic ray induced gamma-ray emission in clusters, both NGCs will be clearly disfavored when only studying DM annihilation.} 
  	\item Ref.~\cite{pinzke11} summarizes in their table 1 absolute fluxes for different galaxy clusters. For their model BM-K', they find that, above 100 GeV, Virgo yields the highest flux level, followed by Fornax, Perseus and Coma. This is perfectly compatible with our results. However, total fluxes are roughly a factor 40 higher in all cases. The differences are mainly due to different substructure treatments, which lead to rather different substructure boosts, as well as to the inclusion of the Sommerfeld effect, which was not considered here.
  	\item To our knowledge, ref.~\cite{gao11} is, together with the present work, the only one in the literature that compares in some detail dwarfs and clusters in the context of DM annihilation. However, it is hard to compare it with our results, as they refer their fluxes to what it is found in N-body cosmological simulations. Qualitatively speaking, both works agree on the fact that Willman~1 is brighter than Draco, and Fornax more promising than Coma. On the other hand, we remind that we find substantially lower substructure boost factors (a factor $\sim$20), and therefore we do not find clusters to be best targets than dwarfs {\it always}, as they do. In our case, the exact ranking is indeed rather sensible to the set of parameters  that we select in order to plan the observational search strategy (see section~\ref{sec:comparison}). 
	\end{itemize}
\end{itemize}

\subsection{Milky Way foreground}  \label{sec:MWforeground}

Yet, we did not take into account the apparent position of the objects with respect to the Galactic Center. This issue might turn out to be relevant, as we should expect an important contribution to the gamma-ray flux coming from DM annihilations in the MW halo, which in this sense might act as a foreground. Nevertheless, the DM annihilation coming from the MW halo should not be considered in this way. Indeed, strictly speaking, it should be treated as an additional signal that would sum up to the expected signal from each of the objects in the sample. Note also that both signals would have the same energy spectrum.

\begin{figure}[t]
\centering
\includegraphics*[scale=0.53]{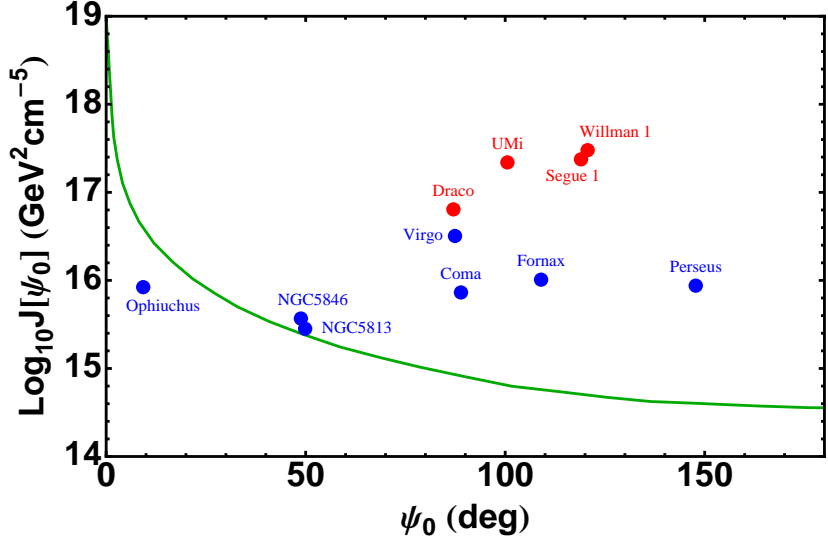}
\caption{DM annihilation flux profile of the MW halo versus the J$_{01}$ values of dwarfs (red points) and clusters (blue points). No substructure boosts are included in the plot. The MW flux profile (green line) was computed assuming a PSF~$= 0.1\grado$. Note that all the objects are above the MW flux profile but Ophiuchus, which surprisingly is well below. The latter will not change by including substructures, as they have a negligible impact in the inner $0.1\grado$.}
\label{fig:MWforeground}
\end{figure}

Figure \ref{fig:MWforeground} shows the DM annihilation flux profile of the MW halo compared to the J$_{01}$ values of both dwarfs (red points) and clusters (blue points) as extracted from tables \ref{tab:dSphs_param} and \ref{tab:clusters_results} respectively. We did not include substructure with the intention to be conservative. The MW flux profile, given by the green line, was computed assuming a PSF~$= 0.1\grado$. Note that all the dwarfs in our sample are well above the MW flux profile. In the case of galaxy clusters, however, only Coma, Fornax, Perseus and Virgo are indeed well above. NGC 5813 and NGC 5846 are just slightly above. And, surprisingly, Ophiuchus is {\it well below} the level of the MW halo, the main reason being the fact that it is only $\sim9\grado$ away from the Galactic Center in the sky. This result will not change even if we had included the effect of substructures, as we remind it has little relevance for the inner $0.1\grado$. Therefore, Ophiuchus is not probably a good candidate after all, as its DM annihilation flux is expected to be completely embedded in the MW foreground (with the PSF used here; see below for further discussion).

On the other hand, from an observational point a view, it might be this MW foreground that better defines the outer extent of the objects in our sample rather than the tidal radii given in table~\ref{tab:dSphs_param} for dwarfs or the virial radius given in table~\ref{tab:clusters_param} for galaxy clusters. It turns out, however, that in case of dwarfs the difference is not very important. Indeed, the level of the DM annihilation from the MW equals the one predicted for dwarfs typically at a radius that {\it already} encloses most of the flux (typically $\sim$95\%). For clusters, however, this issue matters, as clusters have J$_{01}$ values that are substantially lower than those of dwarfs (substructure will not help too much here, as they are not expected to be relevant in the inner 0.1$\grado$). Nevertheless, we note that this effect strongly depends on the integration angle used, as it can been clearly seen in figure 14 of ref.~\cite{charbonnier11} for a sample of dwarfs: the smaller the integration angle the better the contrast between the object signal and the MW foreground. In any case, this fact should be taken into account when programming observational campaigns and analyzing the data. As an example, assuming integration angles of 0.1$\grado$, the MW foreground signal equals the one expected from Perseus without substructure (with substructure) at $\sim$0.2$\grado$ ($\sim$0.55$\grado$) away from its center. At this radius, the integrated signal is roughly 80\% (40\%) of the total flux.

\subsection{Flux predictions} \label{sec:fluxes}

\begin{figure}[t]
\centering
\includegraphics*[scale=0.53]{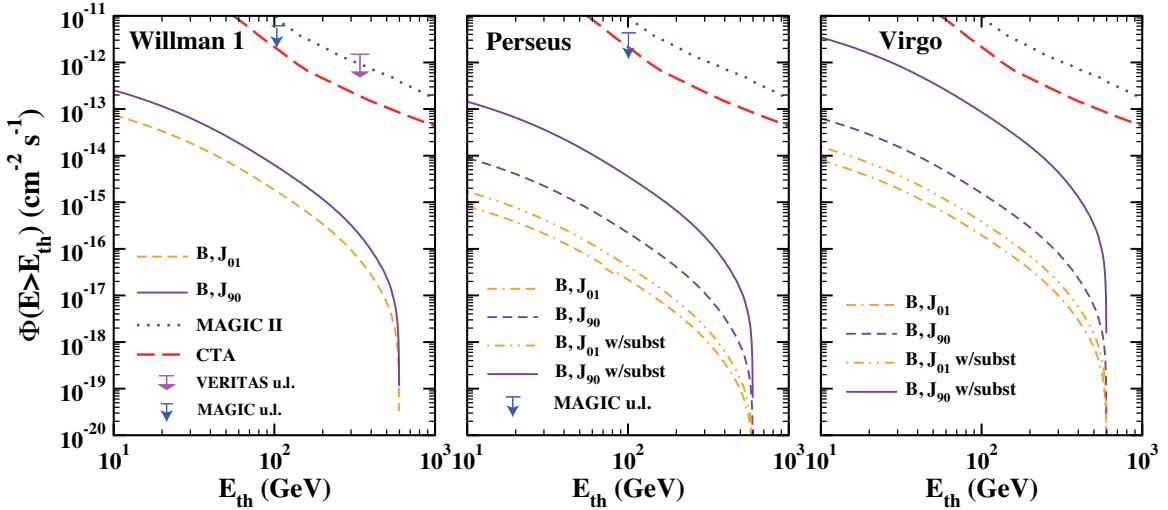}
\caption{
Integrated spectrum of the best dwarf and galaxy cluster in our sample: 
Willman~1 (left panel) and Virgo (right panel), respectively. Given the extremely large spatial extension of the 
flux in Virgo, we also included Perseus (middle panel) as a good alternative (see text for discussion). Two values of the flux are plotted, namely those corresponding to J$_{01}$ 
and J$_{90}$ astrophysical factors as given in tables~\ref{tab:dSphs_results} and 
\ref{tab:clusters_results}. 
For Virgo and Perseus we also show the effect of substructures (table~\ref{tab:clusters_results_sub}). 
We used the particle physics benchmark model 
labelled B in table~\ref{tab} and figure~\ref{fsusydif}, 
which gives the highest particle physics factor.
The integral sensitivity curves of
both the MAGIC telescopes and CTA for 50~h observation time are also shown in the panels, as 
well as the ULs to the flux derived from MAGIC and VERITAS observations of Willman~1 
\cite{MAGICwillman1,veritasdwarfs} and Perseus~\cite{MAGICperseus}.}
\label{fig:detection}
\end{figure}

Figure~\ref{fig:detection} 
summarizes the detection prospects of the best objects in our joint dwarfs + clusters
sample for the MAGIC telescopes and for the future Cherenkov Telescope Array (CTA; 
see Ref.~\cite{ctareport}). More specifically, this Figure shows the integrated spectrum 
of the Willman~1 dwarf galaxy, and Perseus and Virgo galaxy clusters. 
We used the particle physics model labelled  {\it B} in table~\ref{tab}, as this model gives the 
most optimistic results for a large portion of the IACT energy range (see 
figure~\ref{fsusydif}), e.g., $f_{SUSY}=7.2\times$10$^{-33}$ GeV$^2$~cm$^{-5}$ at 100 GeV. 
For the astrophysical factor, we picked the J$_{01}$ 
and J$_{90}$ values given in table~\ref{tab:dSphs_results} for Willman~1, and 
tables~\ref{tab:clusters_results} and~\ref{tab:clusters_results_sub} for Perseus 
and Virgo without and with substructures respectively. 
Virgo, which represents the best cluster 
according to its flux level, is extremely extended; indeed, $\psi_{90}>7^\circ$ in this 
object (see table \ref{tab:clusters_results_sub}), probably meaning a serious handicap 
for IACT observations and data analysis of a possible DM signal. In contrast, 
$\psi_{90} \sim1.2^\circ$ in Perseus, which is also rather similar to other clusters like Ophiuchus or Coma and is
therefore a good representative of all of them (although Ophiuchus seems a slighter better candidate according to tables \ref{tab:clusters_results} and \ref{tab:clusters_results_sub}, we remind that this cluster is probably completely embedded in the MW foreground, which a priori makes it less appealing; see section \ref{sec:MWforeground}).

Willman~1 is the best candidate when neglecting substructures, 
both for J$_{90}$ and J$_{01}$. However, once substructures are properly included, Virgo
yields the highest total flux (well described by J$_{90}$), although Willman~1 still 
remains as the best target regarding J$_{01}$ (meaning more point-like for IACTs, as 
already discussed). Interestingly, Perseus is roughly at the same flux level as
Willman~1 attending to their J$_{90}$, although it is clearly less promising than both Virgo and
Willman~1 according to its J$_{01}$. 

Figure \ref{fig:detection} also shows the integral 
   sensitivity curves of both the MAGIC telescopes and CTA in order to illustrate the DM 
   search potential of both instruments. The MAGIC sensitivity line 
represents the flux needed to reach a 5$\sigma$ significance and $\ge10$ excess event 
in 50 h of data 
\cite{magicsensitivity}. As for CTA, the sensitivity curve corresponds to 50 h observation 
time at a zenith angle of 20$^{\circ}$, as given for configuration $E$ in Ref.~\cite{ctareport}. 
We also included in figure~\ref{fig:detection} the ULs to the central flux as 
derived from MAGIC observations of Willman~1 in Ref.~\cite{MAGICwillman1} using their 
particle physics model $K'$ (which are also rather similar to those deduced from MAGIC 
observations of Segue~1 in Ref.~\cite{MAGICsegue} assuming a spectral index of $-1.5$). 
We also plotted the ULs derived for Perseus from MAGIC data in Ref.~\cite{MAGICperseus}.\footnote{One may wonder why the MAGIC upper limits are better than the MAGIC stereo curve sensitivity. This can be explained by the fact that those ULs are derived for a spectral index of about -1.5, while the sensitivity curves are obtained assuming a Crab Nebula-like spectrum. Also, ULs are always expected to be better than sensitivity curves at parity of conditions.}
 
 A quick comparison 
 between the expected level of the DM annihilation flux and both the sensitivity lines 
 of the MAGIC stereoscopic system and CTA shows that it is unlikely that these instruments
  can detect $\gamma$-rays from DM annihilation in any of these objects (which we recall are
   the most promising among dwarfs and clusters), unless other effects are included that
    may boost the DM signal considerably. Indeed, the minimum difference between the CTA 
    sensitivity line and the DM-induced $\gamma$-ray flux from Virgo, which occurs at
     $\sim$135 GeV, is still larger than an order of magnitude. For Willman~1, the 
     situation is even worse, minimum differences being of the order of $\sim$10$^3$ 
     (note that same factors were recently achieved by the MAGIC collaboration for Segue~1 in Ref.~\cite{MAGICsegue}). 

\begin{figure}
\begin{center}
\includegraphics*[scale=0.55]{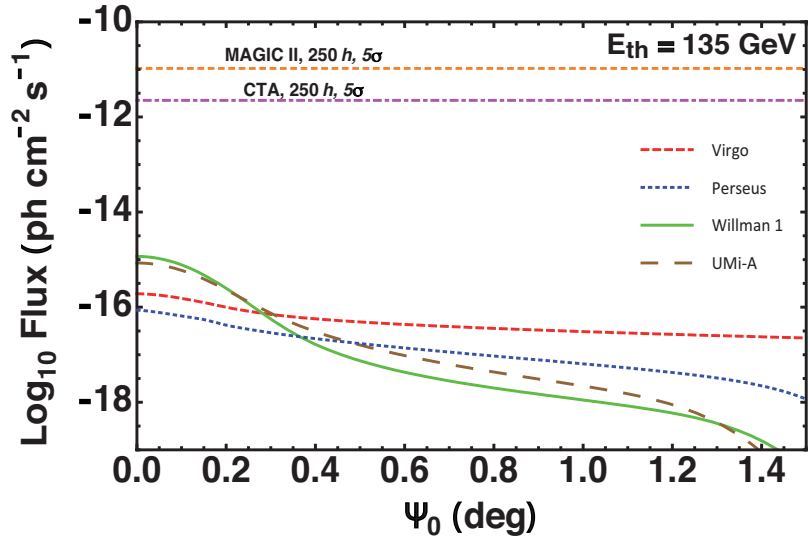}
\end{center}
%\vspace*{0.5 cm}
\caption{DM annihilation flux profiles, including substructures, of the two pairs of
 most promising objects in our sample, namely Willman~1 (green solid line) and Ursa Minor 
 model A (brown dashed) among dwarf galaxies, and Virgo (red short dashed) and Perseus 
 (blue dotted) among galaxy clusters. The sensitivities reached by MAGIC in stereo mode and CTA 
 after a deep observation of 250 h integration time are given for comparison. Both the 
 instrumental sensitivities and the particle physics factor, $f_{SUSY}$, were computed for an 
 energy of 135 GeV, i.e., the one which gives the minimum difference between the expected flux 
 and the sensitivity curves of the MAGIC telescopes and CTA in figure~\ref{fig:detection}.}
\label{fig:flux135}
\end{figure}

At 135 GeV, the value of the particle physics factor, $f_{SUSY}$, is $\sim$0.37$\times$10$^{-32}$ 
GeV$^2$~cm$^{-5}$, and this is indeed the value used in figure~\ref{fig:flux135} to plot the DM 
annihilation flux profiles (with substructure) of Willman~1, Perseus and Virgo together with both 
the MAGIC and CTA sensitivities at 135 GeV. This is therefore one of the most optimistic scenarios that 
it is possible to achieve. We also included in figure~\ref{fig:flux135} Ursa Minor (model A), which is the 
second best dwarf according to our findings in section \ref{sec:dwarfs}. Note that the sensitivity lines 
are now given for a rather optimistic deeper observation of 250 h integration time. However, for the examined particle physics model, the sensitivity lines of current and planned instruments are more than one order of magnitude above the predicted flux profiles.

\section{Summary and conclusions} 	
\label{sec:conclusions}

In this article, we have studied in detail the DM annihilation gamma-ray fluxes for a 
sample of nearby dwarf galaxies (Draco, Ursa Minor, Willman~1, and Segue~1) and nearby 
galaxy clusters (Perseus, Coma, Ophiuchus, Virgo, Fornax, NGC~5813, and NGC~5846), with 
the intention of elucidating which object class (dwarfs or galaxy clusters) is more 
appropriate for gamma-ray DM searches with present-day and/or planned IACTs. We have used the latest modeling of the DM density profiles, which were calculated from the latest observational data available. On the way, we have also discussed some observational and  instrumental aspects that may become crucial when planning a good observational strategy  as well as to posterior interpretation of IACT data. %This was achieved by defining and   calculating some specific quantities that were carefully chosen taking into account both    instrumental and observational properties of IACTs, and that contribute with relevant     information on our understanding of the particular characteristics that a gamma-ray flux     with a DM-annihilating origin may necessarily exhibit. 
We finally studied the detection prospects of the best objects in our sample for current and planned IACTs. The main results of this work can be summarized as follows:
\begin{itemize}
\item 
%Although in the last few years most of the attention in the field of $\gamma$-ray DM searches 
%has been devoted to nearby dwarfs, here we show that 
Nearby galaxy clusters may yield similar, 
or even higher, annihilation fluxes once the effect of substructure in the DM halo is properly taken into account.
\item 
Substructure was included following the 3K10 model \cite{kkk}, 
slightly modified here so that we could safely apply it to CDM halos of different masses rather than MW-sized halos only. We found that substructure is only 
relevant (and critical in most cases) for galaxy clusters and not for dwarf galaxies, indeed 
enhancing the gamma-ray signal a factor $\sim$35-50 in clusters but only 30\% at most in dwarfs.
\item 
Willman~1 appears to be the best candidate among the dwarf galaxies as it shows both the largest 
J$_{90}$ and J$_{01}$, as well as the highest fluxes at the center (see figure~\ref{fig:dwarfprofiles}). 
Moreover, 90\% of the flux is expected to come from a quite small region, i.e., $\psi_{90}\sim0.3$ 
degrees, which means a quite compact emitting region, more feasibly observable with present 
IACTs. Yet, it is not clear whether or not this object is gravitationally 
bound at all, so other candidates with less mass modeling 
uncertainties and quite similar fluxes, such as Ursa Minor and Segue~1, would probably represent better options. 
\item 
Virgo represents the galaxy cluster with the highest fluxes both according to 
J$_{90}$ and J$_{01}$ (essentially due to its extreme proximity). However, its large spatial 
extension (indeed, $\psi_{90} > 7^\circ$ with substructure) can be a serious handicap for IACT 
observations and data analysis. Yet, other candidates with high predicted fluxes and more moderate $\psi_{90}$ values such as 
Perseus or Coma may represent good alternatives.\footnote{Fornax is the best candidate according to ref.~\cite{pinzke11}, as they find this cluster to have a particularly low cosmic ray induced background; however, its angular size $\sim$3$\grado$ is roughly double that of Perseus or Coma, which still makes it rather challenging.}
\item
For an integration angle of 0.1$\grado$, the DM annihilation flux level of the MW halo is well below the one predicted for the inner 0.1$\grado$ of all dwarfs in the sample. This is also true for all galaxy clusters selected but Ophiuchus, located only $\sim$9$\grado$ from the Galactic Center. This cluster appears to be completely embedded in the MW foreground, making it challenging for DM searches. The inclusion of substructures will not affect this result. In any case, we recall that the exact contrast against the MW foreground strongly depends on the integrated angle used.
\item 
The best targets according to J$_{90}$ do not necessarily represent the best targets according to
 J$_{01}$. An example is Segue~1, which is only the fourth best object according to its J$_{90}$ but 
  turns out to be the second best option according to J$_{01}$. This result is due to the interplay in the astrophysical factor  
between the object's distance, the DM distribution and the telescope PSF.
%\item 
%For a given object, the uncertainties in the level of the DM flux introduced by our lack of knowledge 
%of the exact DM density profile are probably only  a factor of a few at $\psi=0^\circ$ and no more
% than an order of magnitude regarding J$_{01}$ (see figure~3 and related discussion in the text). Again, the role of the PSF, which tends to smear out any peculiar feature of the DM annihilation flux profiles, is critical in this result. 
% The importance of these uncertainties contrasts with those expected from the particle 
% physics side, where the allowed models lead to $f_{SUSY}$ values that can span  several orders 
% of magnitude.
\item 
The best dwarf galaxies in the sample have DM annihilation fluxes at $\Psi_0=0^\circ$ which are 
roughly an order of magnitude higher than those expected for the best galaxy clusters. Even when
 including substructure in both kinds of objects, dwarf galaxy flux profiles are higher 
than those of galaxy clusters typically within the inner $\sim0.4^\circ$.
\item 
Once substructures are included, galaxy clusters flux profiles appear to be systematically higher 
than those of dwarfs typically at angles $\gtrsim0.4^\circ$ (right panel in figure~\ref{fig:clustersVSdwarfs}). 
Indeed, the larger the l.o.s.\ angle $\Psi_0$ the larger the relative DM annihilation flux enhancement due to 
presence of substructure. As a consequence, the flux profiles surprisingly remain almost flat up to larger 
radii (of the order of $\psi_{90}$). The fraction of the total DM flux within r$_s$ for 
galaxy clusters, J$_{r_s}$, now drastically decreases from $\sim$0.9 to $\sim$0.2.% (assuming NFW DM density profiles).  
\item 
With the above given considerations in mind, dwarf galaxies are best suited for observational 
strategies based on the search for point-like sources, given their highest J$_{01}$ values and still 
reasonable $\psi_{90}$ values, while galaxy clusters are the best targets for analyses that can deal 
with rather extended emissions (we recall, however, that neither dwarfs nor clusters are point-like 
according to their J$_{01}$/J$_T$ values). Ideally, the regions to be analyzed in clusters should enclose 
at least the solid angle subtended by $\psi_{90}$, of the order of $\sim1^\circ$. Given their 
typical flux profiles, which remain almost flat up to large radii, a suitable data analysis strategy could 
consist of masking those inner regions where other conventional $\gamma$-ray emission mechanisms may 
contaminate the DM signal.
\item 
In the framework of the CMSSM the most optimistic $f_{SUSY}$ value that it is possible to achieve
above 100 GeV is of the order of 10$^{-32}$ GeV$^2$~cm$^{-5}$. The level of the DM annihilation fluxes for the best objects in the joint dwarfs + clusters sample is 
well below the sensitivities of both current IACTs and the future CTA. Indeed, we do find minimum 
differences between predictions and sensitivity lines of more than an order of magnitude in the most optimistic case, i.e., Virgo with 
substructures at 135 GeV (see figure~\ref{fig:detection}) and 50~h observation time of CTA. For 
Willman~1, the mismatch is above two orders of magnitude. Increasing, for example, the total IACT exposure 
 time five 
 times up to 250 h does not change the detection prospects substantially, 
 as the signal-to-noise ratio increases only as the square root of time and therefore both an integrated 
 signal (the one associated either with J$_{01}$ or J$_{90}$, for instance) and the flux profiles are still 
 far from detection.\footnote{Just before the submission of this article, a study of the DM annihilation in classical dSphs 
appeared~\cite{charbonnier11}. 
Although using a different approach and methodology, the authors 
reach rather similar conclusions 
for those objects common to both studies, namely 
Ursa Minor and Draco. In particular, they also find that: i) substructures do not lead to important 
increments of the DM flux in dwarfs; ii) considering the angular extension of the sources is vital 
in order to plan the best search strategy; iii) assuming point-like emission from dSphs is indeed a 
very poor approximation for IACTs, and iv) sensitivities of present and future gamma-ray 
observatories seem to be quite far from detection.}

\end{itemize}

Yet, despite all the negative observational results accumulated so far, as well as the rather discouraging
 detection prospects we found in our analysis, there are reasons for keeping optimistic. First, we remind that our results were achieved assuming a specific particle physics framework, namely the constrained minimal supersymmetric standard model. We chose CMSSM because it represents the benchmark beyond standard model scenario expected to be probed soon (at least part of the parameter space) at the LHC. In any case, as pointed out also in Ref.~\cite{buckley08}, we note that even if the LHC does provide evidence for SUSY or if 
future direct detection experiments will detect a clear signature of DM,  $\gamma$-ray observations 
provide the only way to go beyond a detection in the local DM halo, measuring the DM halo profiles 
and elucidating the exact role of DM in structure formation. 
In other models, there exist new particles and mechanisms that might boost the 
DM-induced $\gamma$-ray signal, e.g., Sommerfeld enhancements~\cite{sommerfeld}. Other 
promising scenarios are also possible which deserve further detailed study, such as DM decay 
\cite{cuesta,ke_decay}. 

New generation IACTs already in operation like the MAGIC stereoscopic system and the HESS upgrade,\footnote{It seems that there is still room for further improvements of the present generation of IACTs, see e.g. Ref.~\cite{becherini11}.} or CTA in the near future, do actually improve the chances of detection significantly and, at least, should be able to impose more stringent limits on the particle physics models.  The complementarity 
of IACT searches with those performed by Fermi/LAT at lower energies also represents a key point in 
order to achieve a more general picture. 
Therefore, $\gamma$-ray DM searches should continue as a top priority for the DM community. Indeed, we are just now 
starting to really unveil the $\gamma$-ray 
energy window, and almost every month new sources are being discovered in the GeV--TeV sky. The time 
for $\gamma$-ray DM searches has definitely come.

\acknowledgments
We are truly grateful to Mike Kuhlen for useful discussions and comments on the implementation
 of the 3K10 substructure model. We also thank Beth Willman and Marla Geha for sharing their thoughts
  and concerns on Willman~1 and Segue~1. We greatly appreciate the help of Juan Betancort, Dan Coe, 
  Michele Doro, Jorge P\'erez-Prieto and Irene Puerto-Gim\'enez. M.A.S.C. also acknowledges the 
  hospitality of Universidad de Huelva, where part of this work was done. M.~C.~is a MultiDark fellow: 
  the authors thank the support of the spanish MICINN Consolider-Ingenio 2010 Programme under grant MultiDark CSD2009-00064.
M.E.G. and M.C. also acknowledge support from the project P07FQM02962 funded by "Junta de Andalucia", 
the Spanish MICINN-INFN(PG21) projects FPA2009-10773 and FPA2008-04063-E.

\end{document}